\documentclass[12pt]{article}
\usepackage{appendix}
\usepackage{anysize}
\usepackage{graphics}
\usepackage{color}
\usepackage{amssymb}
\usepackage{graphicx}
\usepackage{epstopdf}
\usepackage[hang,small,bf]{caption}
\usepackage{hyperref}
\usepackage{amsmath}
\usepackage{latexsym}
\usepackage{setspace}
\usepackage{sectsty}
\usepackage{breqn}

\usepackage{graphicx}
\usepackage{dcolumn}
\usepackage{bm}
\usepackage{hyperref}
\usepackage{epstopdf}
\usepackage{color}

\def\lsim{\raisebox{-4pt}{$\,\stackrel{\textstyle{<}}{\sim}\,$}}
\def\gsim{\raisebox{-4pt}{$\,\stackrel{\textstyle{>}}{\sim}\,$}}

\def\beq{\begin{equation}}
\def\eeq{\end{equation}}
\def\bea{\begin{eqnarray}}
\def\eea{\end{eqnarray}}
\def\ben{\begin{enumerate}}
\def\een{\end{enumerate}}

\def\lsim{\mathrel{\raise.3ex\hbox{$<$\kern-.75em\lower1ex\hbox{$\sim$}}}}
\def\gsim{\mathrel{\raise.3ex\hbox{$>$\kern-.75em\lower1ex\hbox{$\sim$}}}}
\def\ifmath#1{\relax\ifmmode #1\else $#1$\fi}

\def\simlt{\stackrel{<}{{}_\sim}}
\def\simgt{\stackrel{>}{{}_\sim}}

\def \beq{\begin{equation}}
\def \eeq{\end{equation}}
\def \bea{\begin{align}}
\def \eea{\end{align}}

\def\lsim{\mathrel{\rlap{\lower4pt\hbox{\hskip1pt$\sim$}}
    \raise1pt\hbox{$<$}}}                
\def\gsim{\mathrel{\rlap{\lower4pt\hbox{\hskip1pt$\sim$}}
    \raise1pt\hbox{$>$}}}                

\sectionfont{\large\bf}
\subsectionfont{\normalsize\bf}
\subsubsectionfont{\small\bf}
\title{
\vspace*{-1.3cm}
\begin{flushright}
\normalsize{
ANL-HEP-PR-11-58\\
EFI-11-26\\
FERMILAB-PUB-11-477-T}
\end{flushright}
\vspace{0.5cm}
\Large
\textbf{Light Dark Matter and the Electroweak Phase Transition in the NMSSM}
\author{\textbf{Marcela Carena$^{a,b}$, }\\
\textbf{Nausheen R.~Shah$^{a}$, and Carlos E.~M.~Wagner$^{b,c,d}$}\\
[1.5cm]
\normalsize\emph{$^a$~Fermi National Accelerator Laboratory, P.~O.~Box 500, Batavia, IL 60510, USA\footnote{http://theory.fnal.gov}}\\
\normalsize\emph{$^b$~Enrico Fermi Institute and $^c$~Kavli Institute for Cosmological Physics,}\\
\normalsize\emph{University of Chicago, Chicago, IL 60637, USA} \\
\normalsize\emph{$^d$~HEP Division, Argonne National Laboratory, 9700 Cass Ave., Argonne, IL 60439, USA}}}
\begin{document}
\nocite{*}
\setcounter{page}{0}
\maketitle
\thispagestyle{empty}
\vspace{-0.5cm}
\begin{abstract}

We analyze the stability of the vacuum and the electroweak phase transition in the NMSSM close to the Peccei-Quinn symmetry limit. This limit contains light Dark Matter~(DM) particles with a mass significantly smaller than the weak scale and also light CP-even and CP-odd Higgs bosons. Such light particles lead to a consistent relic density and facilitate a large spin-independent direct DM detection cross section, that may accommodate the recently reported signatures at the DAMA and CoGeNT experiments. Studying the one-loop effective potential at finite temperature, we show that when the lightest CP-even Higgs mass is of the order of a few GeV, the electroweak phase transition tends to become first order and strong. The inverse relationship between the direct-detection cross-section and the lightest CP-even Higgs mass implies that a cross-section of the order of 10$^{-41}$ cm$^2$ is correlated with a strong first order phase transition.
\end{abstract}

\newpage

\setcounter{page}{1}

\section{Introduction}

The Standard Model~(SM) agrees remarkably well with all the experimental observables measured presently. However, the experimental tests of the SM have so far been limited to the gauge sector. Far less is known about the Higgs sector responsible for the generation of masses of all SM particles. The smallness of the electroweak symmetry breaking~(EWSB) scale compared to the Planck scale suggests new physics at the weak scale. Among the different possibilities that have been studied in the literature, supersymmetry~(SUSY) is one of the most compelling ones~\cite{reviews}. In a supersymmetric framework the stability of the Higgs mass parameter under quantum corrections can be ensured. In minimal extensions, the SM-like Higgs is naturally light~\cite{mhiggsRG1a}--\cite{Degrassi:2002fi}, and the corrections to electroweak precision and flavor observables tend to be small, leading to good agreement with observations. Additionally, low energy supersymmetry leads to the unification of couplings at large energies and provides a natural DM candidate, namely the lightest neutralino. Moreover, supersymmetric extensions also allows the implementation of the mechanism of electroweak baryogenesis for the generation of the matter-antimatter asymmetry~\cite{Quiros}--\cite{Carena:2008vj}.

The Next-to-Minimal Supersymmetric extension of the Standard Model~(NMSSM)~\cite{Ellwanger:2009dp} preserves all of the above properties, while additionally containing a richer Higgs and neutralino spectrum. This may have an important impact on low energy observables. In particular, if the lighter neutralinos and neutral Higgs bosons are mainly singlets, they would be predominantly produced in association with heavier Higgs bosons or from the cascade decay of other supersymmetric particles, and therefore can easily avoid current experimental constraints.

Recently the DAMA, CoGeNT and CRESST experiments~\cite{Bernabei:2008yi, Aalseth:2010vx, Hooper:2010uy, TAUPDAMA, TAUPCOGENT, TAUPCRESST} have claimed signatures that may be consistent with light DM particles with a mass of about 10~GeV and cross sections of about a few times $10^{-41}$~cm$^2$.  If this particle is identified with a light neutralino, the large cross section necessary to achieve compatibility with these experiments in the MSSM can only be obtained in the very large $\tan\beta$ limit and relatively light CP-odd Higgs mass, $m_A$. However, this region of parameters is severely constrained by non-standard Higgs searches and flavor physics experiments, and is therefore disfavored~\cite{Kuflik:2010ah}--\cite{Cumberbatch:2011jp}.

The NMSSM also provides the possibility of a light neutralino. However, unlike the MSSM case, this neutralino is mostly singlino-like, with a direct dark matter detection cross section mediated predominantly via the lightest CP-even scalar, which is mainly related to the real component of the additional singlet. The proper relic density may also be obtained, with the annihilation of the neutralino dominated by either the light CP-even or the CP-odd Higgs boson, the later also having a large singlet component. In the NMSSM, the light spectrum necessary to explain the above DM is highly constrained in generic regions of parameter space~\cite{Belikov:2010yi}--\cite{Cumberbatch:2011jp}, but can be obtained in the near Peccei-Quinn~(PQ) symmetry limit~\cite{Ciafaloni}--\cite{Draper:2010ew}. Masses of the lightest CP-even Higgs boson of about a few~GeV lead to a large spin independent direct dark matter detection cross section, $\sigma_{SI}$, while the proper relic density may be obtained via close to resonant annihilation mediated by the CP-odd Higgs. The model was shown to be consistent with all experimental constraints, and provides a new paradigm for the study of the NMSSM phenomenology~\cite{Draper:2010ew}.

With these considerations in mind, it becomes important to understand if the mechanism of electroweak baryogenesis may be accommodated within this framework. In this work, we do not analyze the CP-violating sources~\cite{Carena:1997gx}--\cite{Cirigliano:2006dg},\cite{Huber:2006wf}, instead, we concentrate on the conditions necessary for the preservation of the baryon asymmetry, associated with a strong first order phase transition~\cite{Quiros}. Previous works have shown that large values of the trilinear Higgs coupling in the NMSSM tend to induce a strong phase transition~\cite{Pietroni:1992in}--\cite{Profumo:2007wc}. However, for sufficiently low values of the singlet mass parameter, instabilities in the effective potential seem to occur that tend to prevent the successful realization of this scenario~\cite{Menon:2004wv}.

In this article, we show that near the PQ-symmetry limit of the NMSSM, a global physical minimum may naturally be found for either relatively heavy or light singlets. Moreover, in the light singlet case, one obtains a small lightest CP-even Higgs mass which leads to a strong first order phase transition, providing a correlation between large dark matter direct detection cross section and a strong first order phase transition. Light SM-like CP-even Higgs bosons, are helpful in strengthening the first order phase transition and in avoiding possible instabilities.

The article is organized as follows: In section 2 we discuss the properties of the NMSSM near the PQ symmetry limit. In section 3 we look at the EWSB vacuum and the electroweak phase transition necessary for baryogenesis. We further study the EWSB and the phase transition in a Simplified Model and demonstrate that a range of small soft supersymmetry breaking masses for the singlet field leads to both EWSB and a sufficiently strong first order phase transition to the physical vacuum, preserving the baryon asymmetry. In section 4 we present a numerical study of the full model. In section 5 we discuss the phenomenological consequences. We reserve section 6 for our conclusions.

\section{The near Peccei-Quinn symmetry limit of the NMSSM}

We concentrate on the standard NMSSM framework, with a Higgs super-potential
\begin{equation}
W =\lambda  \mathbf{S H_u H_d} + \frac{1}{3}\kappa \mathbf{S}^3.
\end{equation}
The neutral Higgs low-energy effective potential contains the following dominant components
\begin{eqnarray}
V_{\it H} & = & m_{H_u}^2 |H_u|^2  + m_{H_d}^2 |H_d|^2+  \lambda^2 |H_u H_d|^2  +  ~\frac{\left(g_1^2 + g_2^2\right)}{8} \left( H_u^2 - H_d^2\right)^2.
\label{higgspot}
\end{eqnarray}
The singlet dependent scalar potential terms are given by
\begin{eqnarray}
 V_{\it S} &=& {m^2_s}|S|^2   + \lambda^2 |S|^2  \left(|H_u|^2 +| H_d|^2 \right) + \kappa^2 |S|^4 \nonumber\\
&& +\left(  \kappa  \lambda  S^2 H_u^* H_d^* - \lambda A_{\lambda} H_u H_d S+\frac{1}{3} \kappa A_{\kappa} S^3  + h.c \right).
\label{Spot}
\end{eqnarray}
Here $H_d$, $H_u$ and $S$ denote the neutral Higgs bosons corresponding to $\mathbf{H_d}$, $\mathbf{H_u}$ and $\mathbf{S}$ respectively. Following reference~\cite{Draper:2010ew}, we concentrate on the near PQ symmetry limit of this model, namely $\kappa\ll\lambda \lesssim  0.2$. For $\kappa=0$, the PQ symmetry is realized.

A small $\kappa$ denotes an explicit breaking of the PQ symmetry, otherwise spontaneously broken by the presence of a singlet vacuum expectation value~(vev), $\langle S\rangle =S_0\simeq\mu/\lambda$. Hence the CP-odd scalar, $a_1$, behaves as a pseudo Goldstone boson, which consequently acquires a small mass
\begin{equation}\label{ma1}
m^2_{a_1}\simeq -3 \frac{ \kappa }{\lambda} A_{\kappa} \mu\;.
\end{equation}
The phenomenology of a light $a_1$ has been thoroughly studied in the R-symmetry limit~\cite{Dobrescu:2000jt,Dermisek:2005ar}.

A light neutralino also appears in this region of parameters making a clear distinction with the scenario studied previously in Refs.~\cite{Dobrescu:2000jt,Dermisek:2005ar}. The lightest neutralino is mainly singlino, with a mass given by
\begin{equation}\label{mchi1}
m_{\chi_1}\simeq \lambda^2 \frac{v^2}{\mu} \sin2\beta + 2 \frac{\kappa}{\lambda} \mu,
\end{equation}
where $v=174$ GeV and $\tan\beta\equiv\langle H_u\rangle/\langle H_d\rangle$. As was pointed out in Ref.~\cite{Draper:2010ew}, for $\lambda\lesssim 0.2$, $\tan\beta \sim 10$, $\mu\sim $~few hundred GeV and $\kappa/\lambda$ on the order of a few percent, $m_{\chi_1}$ is of order 10 GeV.

Another important effect is associated with the heavy non-standard Higgs bosons in this scenario. For moderate values of $\tan\beta$, the minimization of the effective potential determines that, near the PQ symmetry limit
\begin{eqnarray}
m_{H_d}^2 & \simeq & A_{\lambda}^2\;,
\label{mhd}\\
A_{\lambda} & \simeq & \mu \tan\beta\;.
\label{minrel}
\end{eqnarray}
Since $\mu$ is of the order of a few hundred GeV, this creates a large hierarchical separation amongst the heaviest CP-even, CP-odd and charged Higgs bosons~(which acquire masses of order of a few~TeV for $\tan\beta$ of order 10), the second lightest CP-even Higgs~(with mass close to 120~GeV), and the lightest CP-even and CP-odd Higgs bosons~(with masses about 10~GeV).

 \subsection{Loop Corrections}

Loop corrections are very important in defining the CP-even spectrum. At large values of $\tan\beta$, they lift the mass of the second lightest CP-even Higgs~(which has SM-like properties) above $m_Z$, an effect that has a logarithmic dependence on the third generation squark spectrum. Specifically looking at the corrections induced to the SM-like Higgs quartic coupling:
\begin{eqnarray}
\Delta V_{H^4} &\simeq&\left\{\frac{\Delta \tilde{\lambda}_{\tilde{t}}}{2}  -\frac{3}{16\pi^2}\frac{m_t^4}{v_u^4}\left[ \log\left(  \frac{H_u^2}{v_u^2}\right) - \frac{3}{2}\right] \right\} H_u^4\;,\label{HiggsLoop}\\
&&\nonumber\\
\Delta \tilde{\lambda}_{\tilde{t}} &=&  \frac{3m_t^4}{8 \pi^2 v^4} \left[\log\left(\frac{m_{\tilde{t}}^2}{m_t^2}\right) + \frac{A_t^2}{m_{\tilde{t}}^2}\left(1 - \frac{A_t^2}{12 m_{\tilde{t}}^2} \right) \right],\label{deltaLambda}
\end{eqnarray}
where $m_{\tilde{t}}$ is the characteristic mass of the scalar partners of the top-quark, $A_t$ is the stop mixing parameter and $m_t=y_t v_u$ is the running top-quark mass at the weak scale. $\Delta \tilde{\lambda}_{\tilde{t}}$, comes from the stop one-loop induced corrections, and the second term in $\Delta V_{H^4}$ comes from the top induced corrections in the DR scheme. We have assumed that the stop supersymmetry breaking masses are larger than the weak scale, so the stop induced field dependent logarithmic term is suppressed. This leads to just an effective correction to the $H_u$ quartic coupling, which depends logarithmically on the stop mass scale.

Hence, the loop-corrected mass of the second lightest~(SM-like) Higgs boson, $m_{h_2}$, for moderate or large values of $\tan\beta$ ($v_u \simeq v$), is given by~\cite{mhiggsRG1a}--\cite{HHH}
\begin{equation}
m_{h_2}^2 \simeq  m_Z^2 + 3 \frac{m_t^4}{4 \pi^2 v^2} \left[\log\left(\frac{m_{\tilde{t}}^2}{m_t^2}\right) + \frac{A_t^2}{m_{\tilde{t}}^2}\left(1 - \frac{A_t^2}{12 m_{\tilde{t}}^2} \right) \right],
\label{mh2st}
\end{equation}
where the first term is the tree-level contribution and the second term proceeds from the stop loop corrections ($2 \Delta \tilde{\lambda}_{\tilde{t}} v^2$). The corrections induced by the mixing with the singlet sector become negligible in the region of parameters under study. In our numerical work, we will also always include the one-loop corrections induced by the gauge sector, which, however, lead to very small corrections to the Higgs mass parameters.

In the effective $S$-scalar potential, the large masses of the non-standard Higgs bosons induce a large correction. In particular, the $S$-quartic coupling dependent component of the potential at the weak scale acquires a contribution which is approximately given by
\begin{equation}
\Delta V_{S^4} \simeq \frac{\lambda^4}{16 \pi^2} S^4 \left( 1 + \frac{A_\lambda^2}{ m_{H_d}^2} \right)^2 \log\left( \frac{m_{H_d}^2}{m_t^2}\right)
- \frac{\lambda^4}{8 \pi^2} S^4 \left[ \log\left(\frac{\lambda^2 S^2}{m_t^2}\right) -\frac{3}{2} \right].
\end{equation}
This implies that in the PQ-limit, the dominant contribution to the $S$-quartic coupling may come from loop-corrections, rather than from tree-level contributions. This property has important consequences for the CP-even Higgs spectrum. Defining the small parameter,
\begin{equation}
\varepsilon\equiv \frac{\lambda \mu}{m_{h_2}} \left( \frac{A_\lambda}{ \mu\tan\beta} -1\right)  \sim \mathcal{O}(10^{-2})\;, \label{ourlimit}
\end{equation}
it is easy to show that the lightest CP-even scalar acquires a mass
\begin{equation}
\label{mh1}
m_{h_1}^2  \simeq  -4v^2\varepsilon^2 +\frac{4v^2\lambda^2}{\tan^2\beta} + \frac{ \kappa }{\lambda} A_\kappa \mu + ~4 \frac{\kappa^2}{\lambda^2} \mu^2 + \frac{\lambda^2 \mu^2}{2 \pi^2} \log\left(\frac{\mu^2}{m_t^2}\tan^4\beta \right),
\end{equation}
where the last term proceeds from the one-loop corrections.

\section{Vacuum Stability and Phase Transition}
\subsection{Zero Temperature Properties}

The vacuum at $T=0$ needs to be analyzed to see whether this model gives rise to a realistic EWSB scenario. Hence we need to study the structure of the effective potential given in Eqs.~(\ref{Spot}) and~(\ref{higgspot}) including all relevant one-loop effects:
\begin{equation}\label{VEWSB}
V=V_S+V_H.
\end{equation}

The properties of the NMSSM that we are interested in are controlled by the parameter set: $\{ \lambda, A_\lambda, \kappa, A_\kappa, \tan \beta, \Delta \tilde{\lambda}_{\tilde{t}}\}$, and the soft masses: $\{ m_s^2, m_{ H_d}^2, m_{H_u}^2\}$. The physical requirement of the EWSB extremum to be at $v=\sqrt{|\langle H_u \rangle |^2+|\langle H_d\rangle|^2} \simeq174$~GeV forces $m_{H_u}^2$ and $m_{H_d}^2$ to very specific values for each point in the parameter set. The further requirement that the EWSB extremum must be a minimum and in fact the global minimum, constrains $m_s^2$ to be within a well defined range.

Assuming no spontaneous breakdown of the CP symmetry, the simplest way of ensuring that the EWSB extremum is indeed a minimum is to look at the Hessian of the effective potential, namely the CP-even Higgs mass matrix, ensuring that all the eigenvalues are positive,
\begin{equation}\label{hess}
M_H^{i j}= \frac{1}{2} \frac{\partial^2 V}{\partial x_i \partial x_j},
\end{equation}
where $x_i$ represent the real components of the scalar potential and the second derivative must be evaluated at the corresponding extreme values.

We demand that the physical vacuum be associated with the global minimum and compute physical masses from the eigenvalues of the mass matrix, always using the one-loop effective potential. Since all couplings affecting the singlet sector are weak, we do not expect two loop corrections to modify the lightest CP-even and CP-odd Higgs masses in a relevant way. The SM-like Higgs mass, which depends on the third generation squark spectrum at the one-loop level, as shown in Eq.~(\ref{mh2st}), is moderately modified by two-loop corrections induced by QCD and third generation Yukawa couplings~\cite{mhiggsRG1a}--\cite{Degrassi:2002fi}. These corrections amount effectively to a modification of the effective $\Delta \tilde{\lambda}_{\tilde{t}}$ coupling, which we shall fix in order to acquire $h_2$ masses below 130~GeV, corresponding to the maximum value that may be achieved for a third generation squark spectrum of order 1~TeV. We reserve the detailed study of higher-loop effects at zero and finite temperature to a future investigation.

\subsection{Electroweak Phase Transition at Finite Temperature}

The realization of the mechanism of electroweak baryogenesis, needed to generate the matter-antimatter asymmetry, demands the presence of a strong first order electroweak phase transition. In order to determine if such a transition occurs in the NMSSM, we need to study the temperature evolution of the potential.

The requirement of a strong first order phase transition is that $\phi_c >T_c$, where $T_c$ is the temperature at which the potential at the non-trivial minimum is equal to the value at the trivial minimum and $\phi_c$ is the projection of the Higgs vev in the direction of the Higgs doublets~\cite{Quiros}~\cite{Kajantie:1995kf} at that temperature. If a one-step phase transition is required, the finite temperature minimum, $\phi_c$, evolving to the physical minimum ($v=174$ GeV) at $T = 0$ should be the global minimum at the critical temperature, and should develop first as the Universe cools down. If this is not the case, then we face the situation of a non-physical metastable vacuum developing first, where the Universe may be trapped for long periods of time, or a metastable physical vacuum. To analyze either of these situations, one would have to study the tunneling rates between vacua. This is beyond the scope of this work, and hence we will concentrate on regions of parameter space where a one step electroweak phase transition is realized, and the physical vacuum is stable.

In general, the critical temperature, $T_c$, turns out to be a few tens of GeV, and in order to compute the Higgs effective potential at finite temperature, the high temperature expansion is not a priori justified. Therefore, we use the exact temperature dependent one-loop contribution to the potential, which for a particle of mass $m\equiv m(\phi)$ at temperature $T$ is proportional to~\cite{Dolan:1973qd}:
\begin{equation}
\label{tempCorr}
V(T,m, \pm1)=\mp \frac{T^4}{2 \pi^2} \int^\infty_0 x^2 \log (1\pm e^{-\sqrt{ x^2+\frac{m^2}{T^2}}} )dx,
\end{equation}
where $\pm1$ in the argument of $V$ stands for either fermions or bosons.

We include the one-loop finite temperature contribution associated with the top-quark, the gauge bosons and the Higgsino fields, which provide the most important temperature dependent corrections to the effective potential:
\begin{eqnarray}
\label{Vtemp}
V_T &=& ~8 V(T, \lambda S, +1) +12 V(T, \frac{m_t}{v_u} H_u, +1)+ 3 V(T,\frac{m_Z}{v} \sqrt{H_u^2+H_d^2},-1) \nonumber\\
&&+~6 V(T, \frac{m_W}{v} \sqrt{H_u^2+H_d^2},-1).
\end{eqnarray}

Light Higgs bosons and neutralinos have relatively small couplings to the Higgs and singlet fields and therefore their finite temperature contribution is small. Similarly, gauginos give a very small contribution to the potential which we are neglecting in this work. We are assuming that stops are heavy and therefore decouple from the plasma, leaving only a contribution at zero temperature, which, as shown before, is responsible for raising the SM-like Higgs mass above $m_Z$.

\subsection{Electroweak Phase Transition in a Simplified Model}

In order to understand the properties of the electroweak phase transition in the NMSSM close to the PQ symmetry limit, we first analyze a Simplified Model which, as we shall see, carries all the important features associated with the case under study. Hence, we consider the NMSSM potential in the PQ limit~($\kappa=0$), taking into account only the dominant temperature dependent one-loop corrections affecting the Higgs mass parameters, and the trilinear terms induced by the gauge bosons.

The electroweak symmetry breaking and the phase transition properties are mostly dependent on the minima of the potential, therefore we shall replace $S$ by its extreme value as a function of the Higgs boson fields:
\begin{equation}\label{S0}
S_0= \frac{\tilde{a} \phi^2}{ (m_s^2 +\lambda^2 \phi^2)}\;.
\end{equation}
This leads to the expression~\cite{Menon:2004wv}
\begin{equation}\label{Vphi}
V\left(\phi ,m_s, T\right) = m_{\phi}^2 (T) \phi ^2-T E \phi ^3+\frac{1}{2}\tilde{\lambda }\phi ^4-\frac{\tilde{a}^2\phi ^4}{\left(m_s{}^2+\lambda ^2\phi ^2\right)},
\end{equation}
where $\phi=\sqrt{|H_u|^2+|H_d|^2}$ and $\tilde{a}=\lambda A_\lambda \sin (\beta) \cos (\beta)$. $m_{\phi}^2(T)=m_{\phi}^2(0)+cT^2$, and $c$ is determined by the all weak scale particles that couple to the Higgs and contribute to the finite temperature potential. In order to obtain a SM-like Higgs mass in the range 115--130 GeV, $\tilde{\lambda} \sim 0.25$ while the parameter $E$ proceeds from weak gauge boson effects, $E\sim 0.02$ .

The mass, $m_{\phi}^2(T)$, is the effective Higgs mass parameter including one loop finite temperature corrections, and the zero temperature component, $m_{\phi}^2(0)$, may be extracted from the condition of a vanishing first derivative of the Higgs potential at $\phi = v$ and $T = 0$. The first derivative of the Higgs potential is given by
\begin{equation}\label{Vphiprime}
\phi ^2\frac{\partial V}{\partial \phi ^2}=m_{\phi}^2(T) \phi ^2-\frac{3}{2}T E \phi ^3+\tilde{\lambda }\phi ^4-\frac{\tilde{a}^2\phi ^4\left(2 m_s{}^2+\lambda ^2\phi ^2\right)}{\left(m_s{}^2+\lambda ^2\phi ^2\right){}^2}.
\end{equation}

The vanishing of Eq.~(\ref{Vphiprime}), does not ensure that the zero temperature vacuum is a global minimum. For that to happen, one should ensure that the value of the potential at the electroweak-symmetry breaking minimum is at least deeper than the trivial one, namely
\begin{eqnarray}\label{EWSBV}
V\left(\phi=v ,m_s, T=0\right)~=~-\frac{\tilde{\lambda }v^4}{2}+\frac{\tilde{a}^2m_s^2v^4}{\left(m_s^2+\lambda ^2v^2\right)^2}&< &0\nonumber\\
&&\nonumber\\
\implies \qquad \tilde{\lambda }-\frac{2~  \tilde{a}^2 \; m_s^2}{\left(m_s^2+\lambda ^2v^2\right)^2}&>&0
\end{eqnarray}

The above inequality gives a quadratic in $m_s^2$, which is fulfilled for negative values of this mass parameter, while for positive values of $m_s^2$, solutions are obtained when $m_s \equiv \sqrt{m_s^2}$ is in the range

\begin{equation}\label{msEWSB}
m_s\;\;\;\left\{\begin{array}{l}
 >\;\; \frac{\tilde{a}}{\sqrt{2\tilde{\lambda }}}\left(1+\sqrt{1-\frac{2\tilde{\lambda }}{\tilde{a}^2}\lambda ^2v^2} \right) \\
 \\
 <\;\; \frac{\tilde{a}}{\sqrt{2\tilde{\lambda }}}\left(1-\sqrt{1-\frac{2\tilde{\lambda }}{\tilde{a}^2}\lambda ^2v^2} \right)
 \end{array}
 \right. .
\end{equation}

The critical temperature, $T_c$, is defined as the one for which the value of the potential at the non-trivial minimum, $\phi_c\neq0$, is degenerate with the one at $\phi = 0$:
\begin{equation}
 \frac{\partial V\left(\phi ,m_s, T_c\right)}{\partial \phi}~\Biggr\vert_{~\phi=\phi_c}=~ 0\;, \;\;\;\;\;\;\;\;\;
 V\left(0 ,m_s, T_c\right)=
 V\left(\phi_c ,m_s, T_c\right)\;.
\end{equation}
The phase transition strength, which is determined by the ratio of $\phi_c/T_c$, can be obtained by setting both Eqs.~(\ref{Vphi}) and (\ref{Vphiprime}) to zero
\begin{equation}\label{solm}
E T_c\phi _c^3-\tilde{\lambda }\phi _c^4+\frac{2~\tilde{a}^2m_s^2\phi _c^4}{\left(m_s^2+\lambda ^2\phi _c^2\right)^2}=0.
\end{equation}
To ensure a strong phase transition as required for electroweak baryogenesis, one should demand $\phi_c/T_c>1$. Hence, a strong first order phase transition is realized if:
\begin{eqnarray}
\frac{\phi _c}{T_c}~=~\qquad \frac{E}{\tilde{\lambda }-\frac{2~\tilde{a}^2 m_s^2}{\left(m_s^2+\lambda ^2\phi _c^2\right)^2}}\quad& >& \quad 1\label{phiTc}\\
&&\nonumber\\
~\implies  ~ F(m_s^2) ~\equiv ~ \frac{1}{\tilde{\lambda }-\frac{2~\tilde{a}^2m_s^2}{\left(m_s^2+\lambda ^2\phi _c^2\right)^2}}-\frac{1}{E}\; \; &>& \quad 0\;.\label{Fms}
\end{eqnarray}
Note the explicit dependence on $\phi_c$.  The values of $\phi_c$ and $T_c$ may be obtained by using Eq.~(\ref{solm}) and the temperature dependence of $m_{\phi}^2(T)$, namely
\begin{equation}
c \;T_c^2 = G(v) - G(\phi_c) + \frac{3 \phi_c^2}{2} \left(\tilde{\lambda}-\frac{2 \tilde{a}^2 m_s^2}{\left(m_s^2 +\lambda^2 \phi_c^2\right)^2}\right)\;,
\label{phicTc2}
\end{equation}
with
\begin{equation}
G(\phi) =  \left[\tilde{\lambda}  -  \frac{\tilde{a}^2\left(2 m_s^2 + \lambda^2 \phi^2\right)}{\left(m_s^2 + \lambda^2 \phi^2\right)^2}\right]\phi^2\;.
\end{equation}

\begin{figure}[!tb]
\begin{center}
$
\begin{array}{cc}
\qquad (a) & \qquad (b)\\
&\\
\includegraphics[scale=0.4, angle =0]{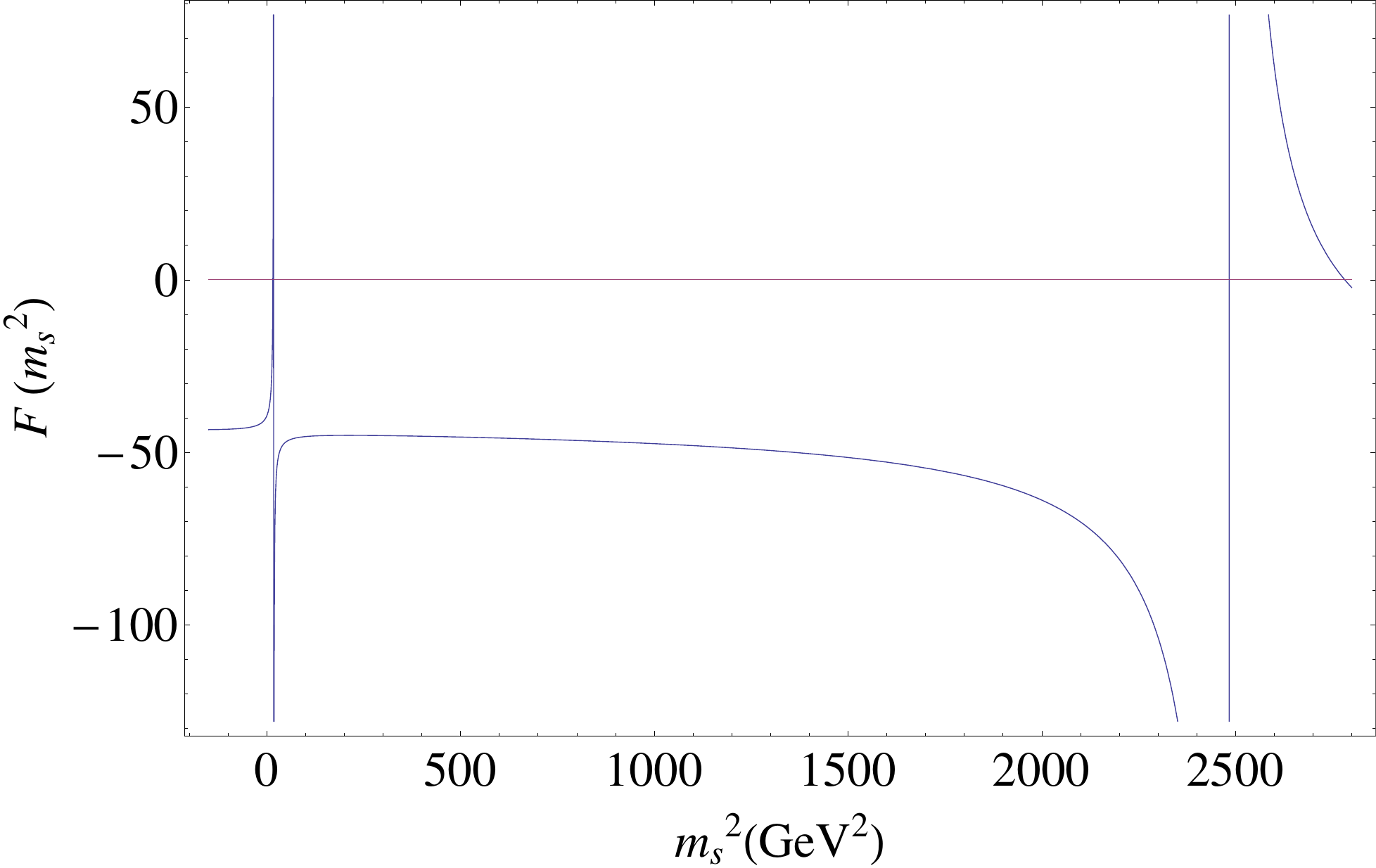}
 &
\includegraphics[scale=0.4, angle=0]{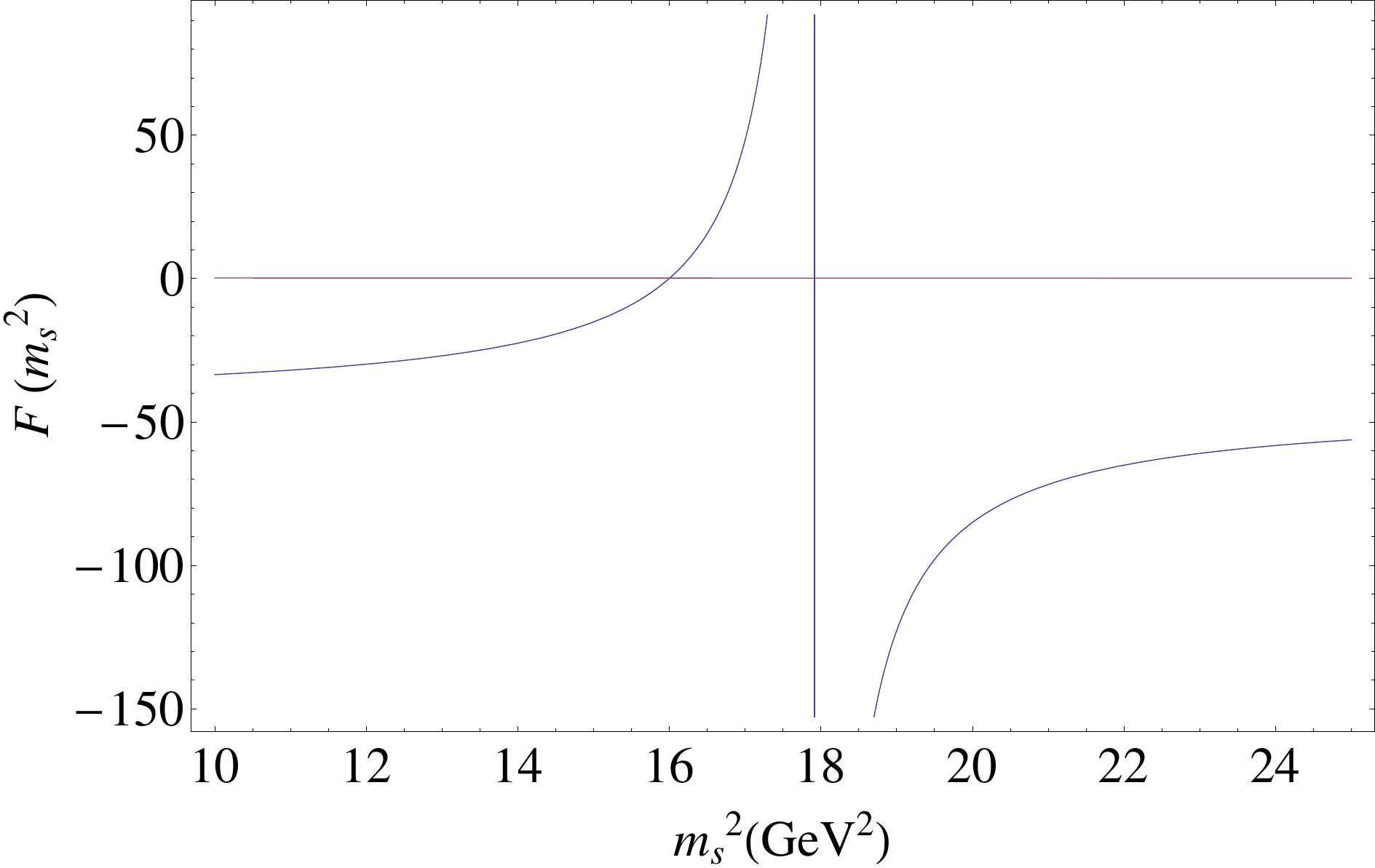}
\\
\end{array}
$
\end{center}
\caption{\footnotesize{$F(m_s^2)$ plotted as a function of $m_s^2$ with $\phi_c=140$ GeV, for Parameter Set 1 in Table~\ref{tablemsAll}. We see that it is only possible to have a strong first-order phase transition~($F(m_s^2) > 0$) for two regions of $m_s^2$. The solution for small $m_s^2$ is shown in more detail in the right plot $(b)$.}}
\label{ms2plots}
\vspace{0.3cm}
\end{figure}

Instead of solving the system, however, we shall concentrate on Eq.~(\ref{Fms}), since it leads to valuable insight into the range of parameters leading to a strong first order phase transition. Similarly to Eq.~(\ref{EWSBV}), Eq.~(\ref{Fms})  leads to a quadratic condition on $m_s^2$:
\begin{equation}
(\tilde{\lambda }-E)-\frac{2~\tilde{a}^2 \; m_s^2}{\left(m_s^2+\lambda ^2\phi_c^2\right)^2}~<~0\;,
\end{equation}
with solutions shifted from those for EWSB, Eq.~(\ref{msEWSB}), by $\tilde{\lambda} \to (\tilde{\lambda}-E)$ and $v \to \phi_c$ and with the inequalities reversed:
\begin{equation}\label{msBar}
m_s \;\;\;\left\{\begin{array}{l}
 <\;\; \frac{\tilde{a}}{\sqrt{2(\tilde{\lambda }-E)}}\left(1+\sqrt{1-\frac{2~(\tilde{\lambda }-E)}{\tilde{a}^2} \lambda ^2\phi_c^2} \right) \\
 \\
 >\;\; \frac{\tilde{a}}{\sqrt{2(\tilde{\lambda }-E)}}\left(1-\sqrt{1-\frac{2~(\tilde{\lambda }-E)}{\tilde{a}^2}\lambda ^2\phi_c^2} \right)
 \end{array}
 \right. .
\end{equation}
However, note that we also need to impose that the denominator in Eq.~(\ref{phiTc}) is positive, as we have implicitly assumed to derive Eq.~(\ref{msBar}). This adds the additional constraint:
\begin{equation}\label{posTphi}
\tilde{\lambda }-\frac{2~\tilde{a}^2 \; m_s^2}{\left(m_s^2+\lambda ^2\phi_c^2\right)^2}~>~0\;.
\end{equation}
Generally $\phi_c$ tends to be smaller than $v$, hence, the above condition gives a stronger constraint than that given by EWSB from Eq.~(\ref{EWSBV}) for $\phi_c<v$. Hence, in this Simplified Model, we see that the only allowed values for small $m_s$ are given in the range:
\begin{equation}\label{msrange}
\;\; \frac{\tilde{a}}{\sqrt{2(\tilde{\lambda }-E)}}\left(1-\sqrt{1-\frac{2(\tilde{\lambda }-E)}{\tilde{a}^2}\lambda ^2\phi_c^2} \right) < m_s < \;\; \frac{\tilde{a}}{\sqrt{2\tilde{\lambda }}}\left(1-\sqrt{1-\frac{2\tilde{\lambda }}{\tilde{a}^2}\lambda ^2\phi_c^2} \right)\;.
\end{equation}

\begin{figure}[!tb]
\begin{center}
\includegraphics[width=0.7\textwidth]{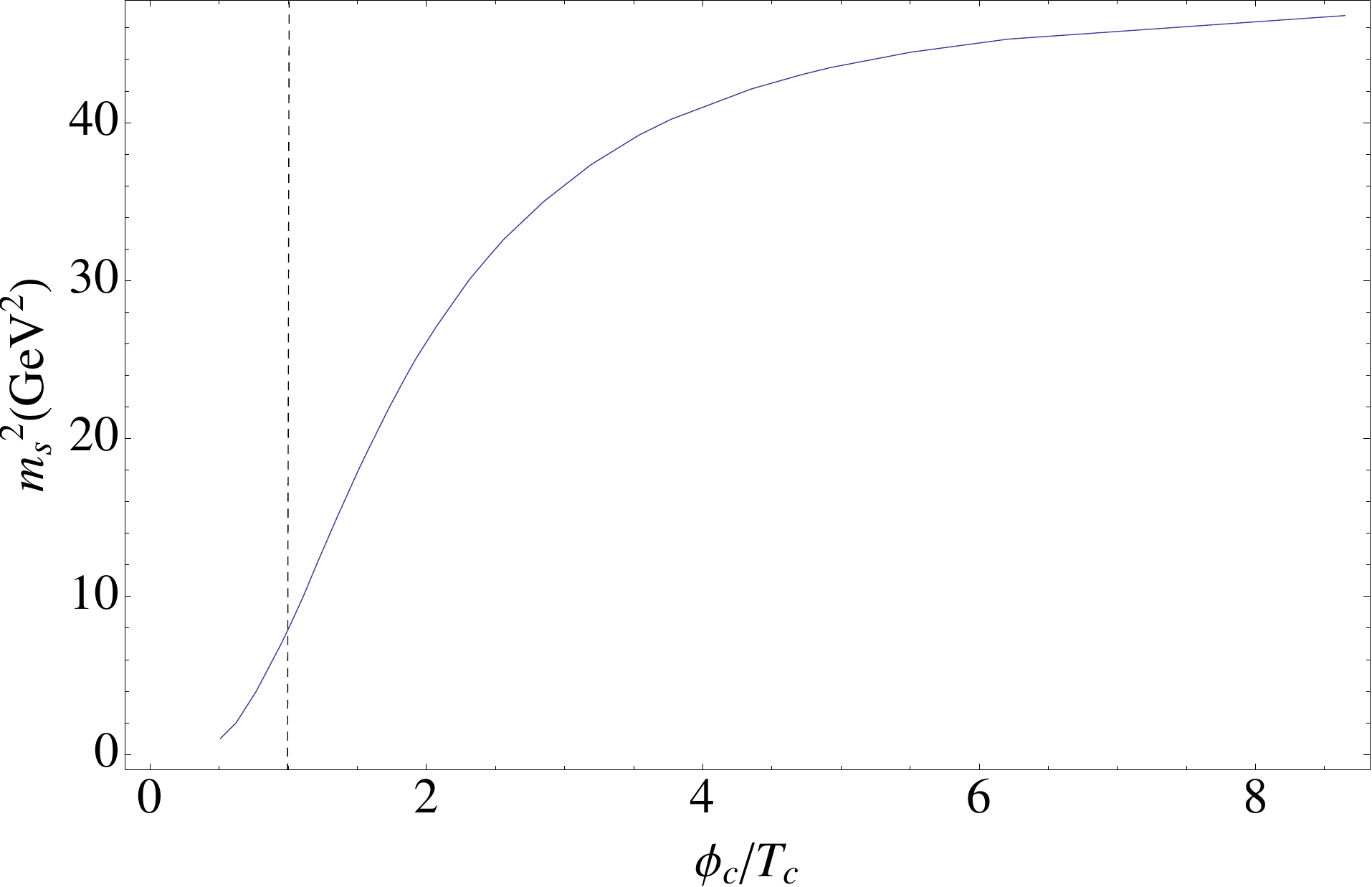}
\caption{\footnotesize{Values of $\phi_c/T_c$ as a function of $m_s^2$ for Parameter Set 1 in Table~\ref{tablemsAll}. }}
\label{simplified1}
\end{center}
\end{figure}

This behavior is clarified if one looks at the behavior of $F(m_s^2)$ for a fixed set of parameters as plotted in Fig.~\ref{ms2plots}. The region enclosed by the two roots in Eq.~(\ref{msBar}) should be the region in Fig.~\ref{ms2plots} where $F(m_s^2)>0$, however, one clearly sees the two poles corresponding to the zeros of the function given in Eq.~(\ref{posTphi}), between which the function again becomes negative and hence violates the required condition. The region of large $m_s^2$ corresponds to relatively large values of the singlet mass, which were studied in Ref.~\cite{Menon:2004wv} and are not consistent with a very light DM particle. The right-hand side of Fig.~\ref{ms2plots} shows the region near the pole and the root corresponding to the region of small $m_s^2$. We clearly see that  for each value of $\phi_c$ there is only a  narrow region of $m_s$ that satisfies both the inequalities given in Eqs.~(\ref{msBar}) and (\ref{posTphi}) and hence is enclosed by the range given in Eq.~(\ref{msrange}).

The bound on $m_s$ derived above has a dependence on $\phi_c$.  Since $\phi_c$ is not independent of $m_s$, what is generally obtained is a  band of values of $m_s$ for which the condition of a strong first order phase transition is satisfied. In Fig.~\ref{simplified1} we present the values of $\phi_c/T_c$ as a funciton of $m_s^2$ obtained by solving the system of equations, Eqs.~(\ref{solm}), (\ref{phicTc2}),  for the same set of parameters as in Fig.~\ref{ms2plots}.   As the phase transition becomes stronger, the value of $\phi_c$ increases towards $v$ and the upper bound on $m_s^2$, Eq.~(\ref{msrange}), moves upwards.   As we shall demonstrate in the next section, a similar upper bound to the one appearing in Eq.~(\ref{msrange}) also appears in the full theory, once the one-loop correction as well as the small $\kappa$ induced contributions are taken into account.  The behavior of $\phi_c/T_c$ in the full theory is also similar to the one depicted in Fig.~\ref{simplified1}.  Finally, the values of $\phi_c/T_c$ turn out to be rather insensitive to the precise value of $E$. A change of $E$ from 0.01 to 0.04 results in a less than ten percent change in $\phi_c/T_c$, and hence, contrary to the SM case, the Debye screening of the gauge boson longitudinal modes has very little effect on the phase transition strength.

For the small $m_s^2$ region, $S_0$, Eq.~(\ref{S0}) (and hence  $m_{h_1}$, Eq.~(\ref{mh1})), decreases for increasing values of $m_s^2$.  As can be observed in Fig.~(\ref{simplified1}), in this region larger values of $m_s^2$  are associated with larger values of $\phi_c/T_c$. Hence,  we conclude a stronger first order phase transition is achieved for  smaller values of $m_{h_1}$.  Finally, let us stress that since generically $\phi_c \simlt v$, large values of $\phi_c/T_c \gg 1$, as the ones achieved for the largest values of $m_s^2$ in Fig.~\ref{simplified1}, may only be obtained for relatively small values of the critical temperature, of order of a few tens of GeV. We shall return to this question in the next section.

\section{Numerical Analysis}

In our numerical study, we first performed scarce scans over parameters somewhat beyond the near PQ symmetry limit, going up to values of $\lambda \simeq 0.5$, $\kappa \simeq 0.01$, $\lambda A_\lambda \simeq 500$~GeV and $\kappa A_\kappa \simeq-0.1$~GeV. 
We selected regions of parameter space for which an electroweak symmetry breaking minimum, $\phi=v$, with moderate values of $\tan\beta$, develops at zero temperature. This corresponds to finding a range of $m_s^2$ for a given set of parameters where all the eigenvalues of the Higgs mass matrix are positive. We then study the temperature evolution of the associated potential. If the phase transition is first order, we determine the critical temperature. In general, the values of $\tan\beta$ at the minimum at $T = T_c$ are slightly different from the ones at zero temperature. Regions where we found a strong first-order phase transition at finite temperature~($T_c< \phi_c$), with a global minimum coinciding with the zero temperature physical minimum, will be referred to as ``Baryogenesis'', since these are the only regions of parameters where the baryon asymmetry may be preserved.

We performed several scans, which were sufficient to identify regions of interest where a strong first order phase transition takes place:
\begin{eqnarray}
\label{ParRange}
\lambda & \simeq & 0.100 \; {\rm to} \;  0.130,
\nonumber\\
\kappa & \simeq & 0.001,\; {\rm to} \;  0.005,
\nonumber\\
\kappa A_{\kappa} & \simeq & -0.1 \;  {\rm to} \;  -0.05,
\nonumber\\
\lambda A_{\lambda} & \simeq & 250 \; {\rm to } \; 450~{\rm GeV} ,\nonumber\\
\tan\beta & \simeq & 10 \; {\rm to} \; 20, \nonumber\\
m_{h_2} & \simeq & 100\; {\rm to } \; 130~{\rm GeV} .
\end{eqnarray}

Observe that we consider values of $m_{h_2}$ smaller than the SM Higgs LEP bound, since, as we shall discuss in the next section, these values may be allowed in this model due to the possibility of non-standard decay modes. On the other hand, we only consider $m_{h_2}\gsim 100$ GeV, since lower values would require lighter stop masses which may be in tension with direct experimental search for these particles, as well as with precision electroweak constraints. As emphasized before, in the region of parameters we consider, the stops are heavy enough to decouple from the plasma at the phase transition temperature.

\begin{table}[!t]
\caption{\footnotesize{Examples of parameter sets of values found where a first order phase transition with $T_c<\phi_c$ takes place. We have fixed the stop spectrum appropriately so that it leads to $m_{h_2} \simeq$115--120~GeV.}}
\begin{center}
\begin{tabular}{@{}|c|c|c|c|c|c|@{}}
\hline
\hline
& $\tan \beta $ & $\lambda$  & $\lambda A_\lambda$ & $\kappa$ & $\kappa A_\kappa$ \\
\hline
\hline
Set 1 & 13 & 0.10375 & 250 & 0.006875 & -0.06 \\
\hline
Set 2 & 13 & 0.10375 & 250 & 0.006875 & -0.08 \\
\hline
Set 3 & 13 & 0.10375 & 350 & 0.001250 & -0.06 \\
\hline
Set 4 & 13 & 0.10375 & 350 & 0.001250 & -0.08 \\
\hline
Set 5 & 18  & 0.10375& 350 & 0.006875 & -0.06 \\
\hline
Set 6 & 18 & 0.10375 & 350  & 0.006875 & -0.08 \\
\hline
Set 7 & 18  & 0.10375 & 450 & 0.001250 & -0.06 \\
\hline
Set 8 & 18 & 0.10375& 450  & 0.001250 & -0.08 \\
\hline
\end{tabular}
\end{center}
\label{tablemsAll}
\vspace{.5cm}
\end{table}

\begin{figure}[!b]
\vspace{0.1cm}
\begin{minipage}[b]{0.49\linewidth}
\centering
\includegraphics[width=\textwidth,trim=.8in 3.5in 0.9in 3.7in,clip=true]{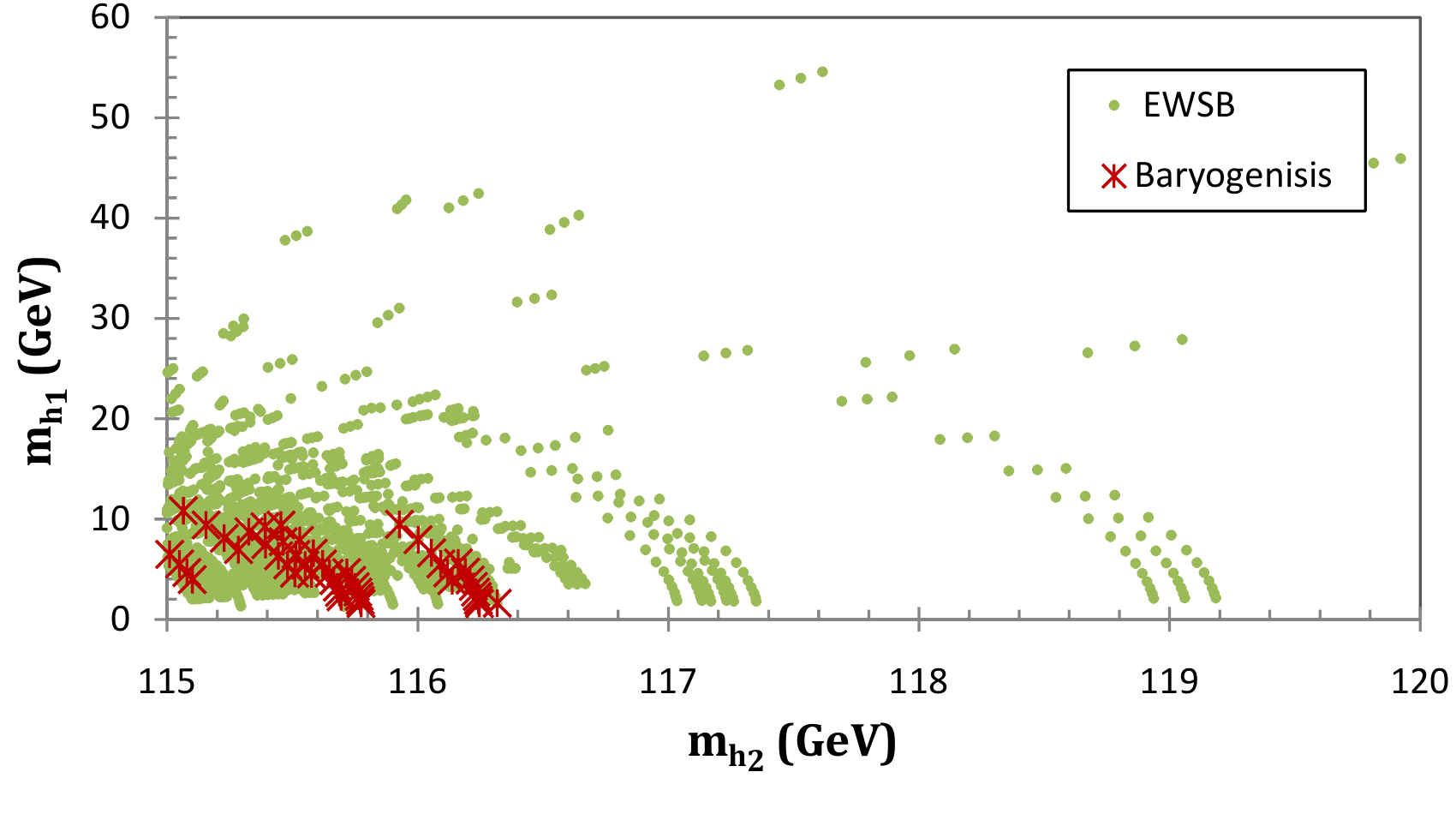}
\end{minipage}
\hspace{0.2cm}
\begin{minipage}[b]{0.49\linewidth}
\centering
\includegraphics[width=\textwidth,trim=.8in 3.5in 0.9in 3.7in,clip=true]{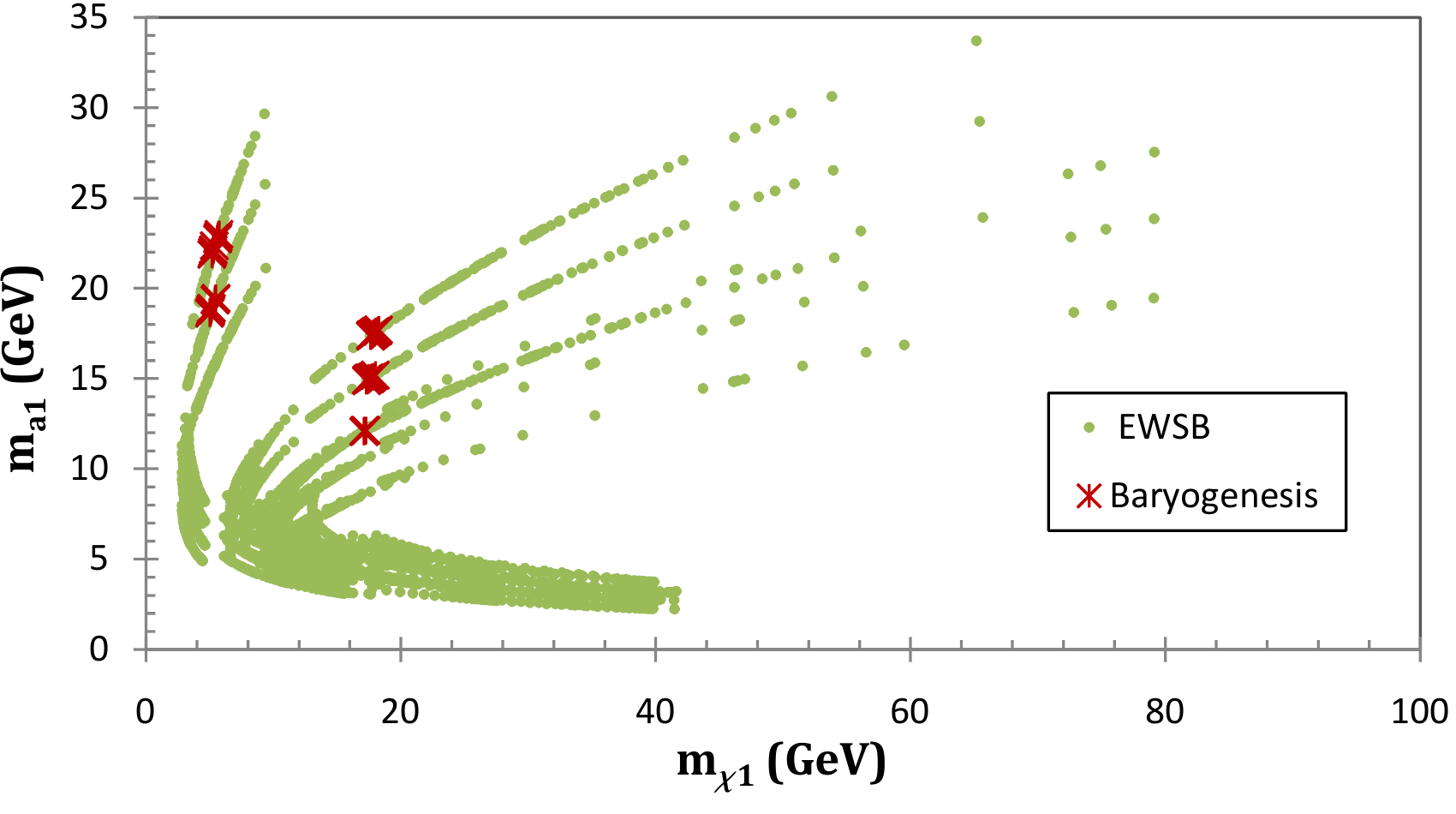}
\end{minipage}
\caption{\footnotesize{\textit{Left:} $m_{h_2}$ vs. $m_{h_1}$ for both EWSB (Green dots) and Baryogenesis (Red crosses), fixing the stop spectrum so that $m_{h_2}$  is in the 115--120 GeV range and scanning over all other parameters. \textit{Right:} Same as left side but for $m_{a1}$ vs. $m_{\chi_1}$. }}
\label{mh1mh2ma1mchiAll}
\vspace{0.5cm}
\end{figure}

Table~\ref{tablemsAll} lists some examples of parameters for baryogenesis found in one of our initial sparse scans, where we found solutions only for relatively small values of $\lambda$ . We show this example for fixed values of $\lambda$ since it allows a better comparison with the Simplified Model. The corresponding relevant masses are presented by red crosses in Fig.~\ref{mh1mh2ma1mchiAll}. The green dots in this figure represent solutions obtained for EWSB that did not lead to baryogenesis. Note that Baryogenesis is only achieved for values of the lightest CP-even Higgs mass below 10~GeV, the CP-odd Higgs mass  below 25 GeV and the neutralino mass  below 20 GeV.

Performing a more thorough scan  around the points in the ``Baryogenesis'' regions, Eq.~(\ref{ParRange}) we were able to find points where the relic density acquires acceptable values. 
Table~\ref{table_msmh} lists two of the parameter sets, $a$ and $b$, for which we were able to find regions consistent with the observed dark matter density. Most of our detailed analysis will be focused on these parameter sets. The stop spectrum was varied such that $m_{h_2}$ takes values in the 100--130 GeV range. The lightest CP-odd Higgs mass remains below 20 GeV, Eq~(\ref{ma1}), and the neutralino mass consistent with the proper relic density tends to be below 10 GeV, Eq.~(\ref{mchi1}).

\begin{table}[!t]
\vspace{0.4cm}
\caption{\footnotesize{Parameter sets of values found where the potential undergoes a strong first order phase transition with $T_c<\phi_c$. We implicitly vary the stop spectrum so that $m_{h_2}$ is scanned over 100 - 130 GeV. We were also able to generate reasonable values for the dark matter density and the spin-independent direct DM detection cross-section. }}
\begin{center}
\begin{tabular}{@{}|c|c|c|c|c|c|@{}}
\hline
\hline
& $\tan \beta $ & $\lambda$  & $\lambda A_\lambda$ & $\kappa$ & $\kappa A_\kappa$ \\
\hline
\hline
Set a & 13 & 0.125& 375  & 0.0021 & -0.055 \\
\hline
Set b & 13 & 0.105& 350  & 0.0023 & -0.080 \\
\hline
\end{tabular}
\end{center}
\label{table_msmh}
\end{table}

\begin{figure}[!b]
\vspace{.1cm}
\begin{minipage}[b]{0.48\linewidth}
\centering
\includegraphics[width=\textwidth,trim=.8in 3.5in 0.95in 3.7in,clip=true]{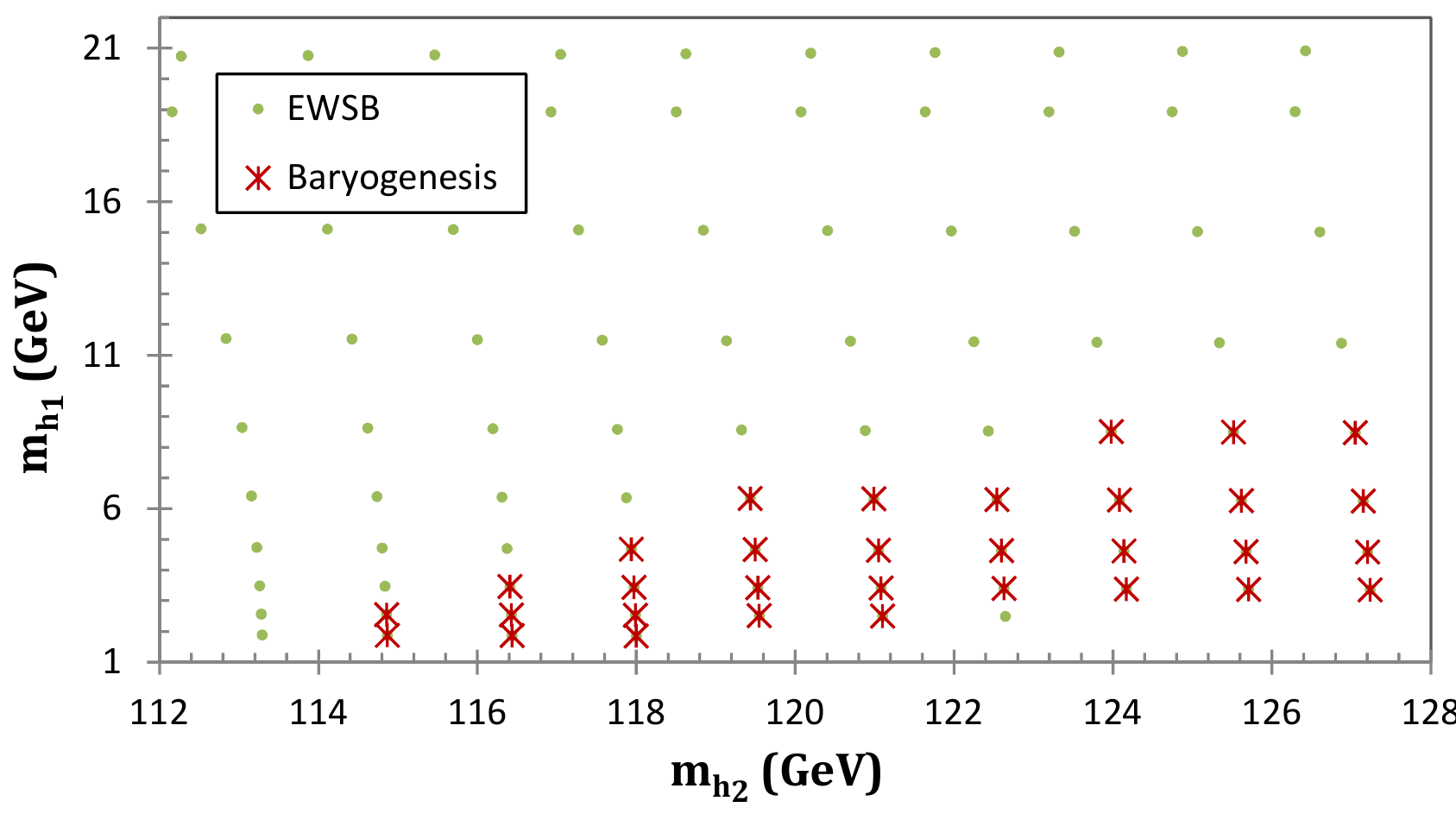}
\end{minipage}
\hspace{0.2cm}
\begin{minipage}[b]{0.48\linewidth}
\centering
\includegraphics[width=\textwidth,trim=.75in 3.5in 1in 3.7in,clip=true]{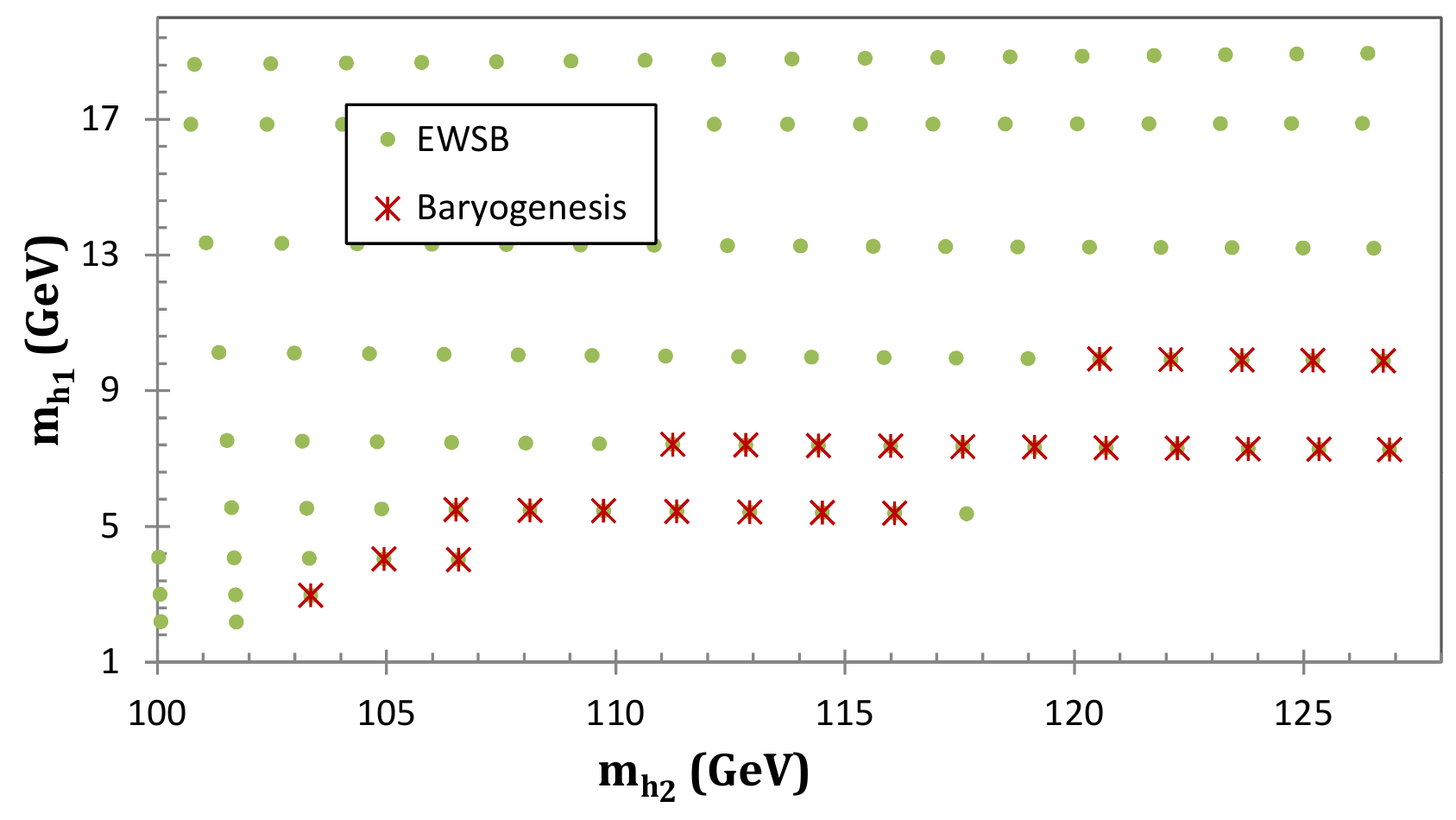}
\end{minipage}
\vspace{0.5cm}
\caption{\footnotesize{$m_{h_2}$ vs. $m_{h_1}$ for both EWSB (Green dots) and Baryogenesis (Red crosses) for Parameter Set $a$ (left) and $b$ (right). }}
\label{mh1mh2ab}
\vspace{0.5cm}
\end{figure}

The masses of the two CP-even Higgs, $m_{h_1}$ and $m_{h_2}$, corresponding to these parameter sets are shown in Fig.~\ref{mh1mh2ab}. As can be explicitly seen in these figures, the solutions with the lowest values of $m_{h_1}$ are associated with the lowest values of $m_{h_2}$. In particular for a given small value of $m_{h_1}$, solutions compatible with EWSB require relatively small values of $m_{h_2}$. The reason for this behavior is that otherwise new minima develop making the electroweak symmetry breaking vacuum metastable. As mentioned before, these additional minima tend to be associated with very small values of $S_0$ and $H_d$. The situation is somewhat complex due to the many parameters involved, however, we can get a qualitative understanding of it by looking at the effective Higgs doublet mass parameter, given by
\begin{equation}
m_{\phi}^2 \simeq m_{H_u}^2 \sin^2\beta + m_{H_d}^2 \cos^2\beta.
\end{equation}
Since $m_{H_d}^2$ is very large, Eq.~(\ref{mhd}), $m_{H_u}^2$ is smaller than $m_{\phi}^2$. The EWSB physical minimum, with the required values of $\tan\beta$, is deeper than the one with $S=H_d =0$, due to the effects induced by the trilinear Higgs coupling through relatively large values of $S_0$ as can be seen from the Simplified Model, Eqs.~(\ref{S0}) and (\ref{Vphi}). As $m_s^2$ increases, the value of $S_0$ becomes smaller and for the same value of $m_{H_u}^2$, the potential at the physical minimum  takes higher values, eventually higher than the ones at $S_0 =H_d = 0$. Hence, if we wanted to keep the physical minimum deeper as we raise $m_s^2$ one would need to raise the value of $m_{H_u}^2$. Since smaller values of $m_{h_1}$ are associated with smaller values of $S_0$, and smaller values of $m_{h_2}$ are associated with larger (or less negative) values of $m_{H_u}^2$, the stability condition is more likely to be fulfilled if $m_{h_2}$ is pushed to lower values for small values of $m_{h_1}$.

Note also that  parameter set $a$ allows for solutions where the mass of the lightest CP-even Higgs, $h_1$, can be very small, of order 1~GeV and the SM-like Higgs mass, $m_{h_2}$, is above the LEP SM-Higgs mass limit. These solutions are very interesting since such small values of $m_{h_1}$ are associated with large values of the spin-independent direct dark matter detection cross section~\cite{Draper:2010ew}, as the ones necessary to explain the intriguing signatures at the DAMA, CoGeNT and CRESST experiments.

\subsection{Comparison with the Simplified Model}

\begin{figure}[!t]
\begin{center}
\includegraphics[width=0.9\textwidth,trim=.8in 3.3in 0.8in 3.5in,clip=true]{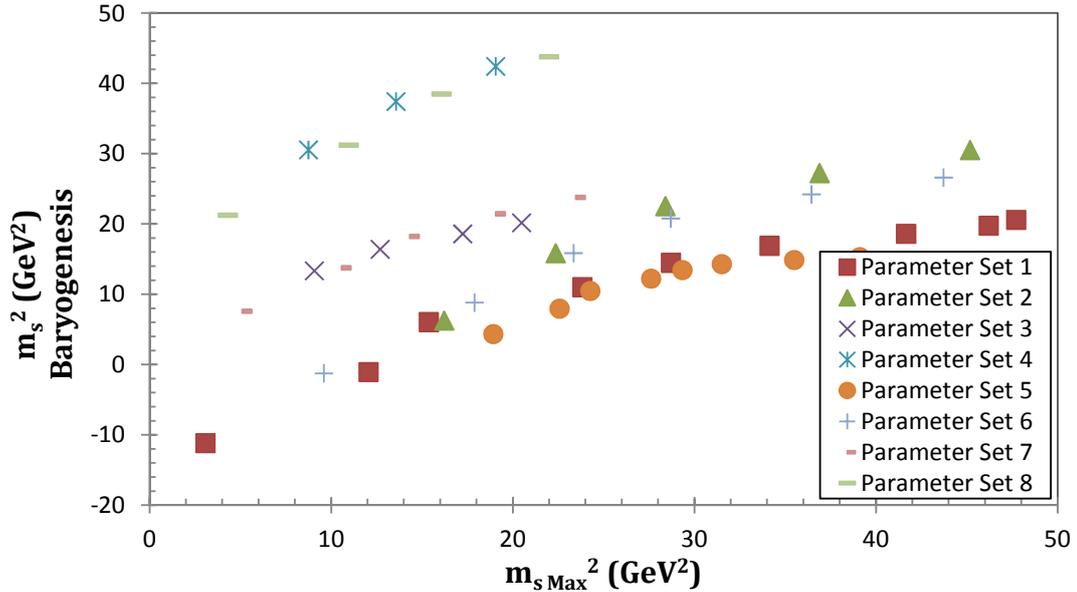}
\caption{\footnotesize{Parametric plot of $m_s^2$ range for set of parameters given in Table~\ref{tablemsAll} and the maximum allowed value approximated in the PQ-limit given in Eq.~(\ref{msrange}). $\Delta\tilde{\lambda}_{\tilde{t}}$ was fixed such that $m_{h_2}$ is in the range 115--120 GeV. }}
\label{ms2limitAll}
\end{center}
\end{figure}

\begin{figure}[!t]
\begin{center}
\includegraphics[width=0.9\textwidth,trim=.9in 3.4in .9in 3.7in,clip=true]{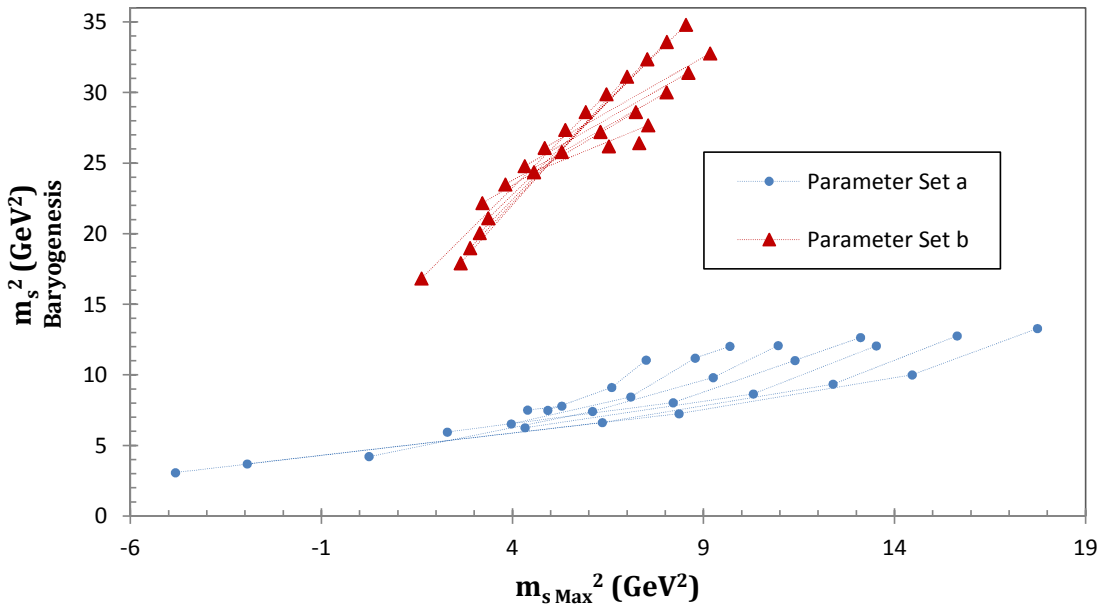}
\caption{\footnotesize{Same as Fig.~\ref{ms2limitAll}, but with $\Delta\tilde{\lambda}_{\tilde{t}}$ scanned such that the $m_{h_2}$ range is $100 -130$ GeV, with fixed parameter sets $a$ and $b$ given in Table~\ref{table_msmh}. Points joined together by dotted line denote same $\Delta\tilde{\lambda}_{\tilde{t}}$.}}
\label{ms2limmh}
\end{center}
\end{figure}
In general we observe that a strong first order phase transition may only be obtained for a small range (band) of $m_s^2$ which depends on the parameters of the model. It is not possible in the full model to write an analytical form for the range as was done in Eq.~(\ref{msrange}). However, we would like to understand whether our insights from the Simplified Model are valid in the full model.

To aid us in understanding the behavior of the potential, we compare the values of $m_s^2$ found for baryogenesis in the full numerical simulation with the upper limit given in Eq.~(\ref{msrange}) for the Simplified Model. Note that we could have compared either of the limits, since the range of $m_s^2$ for a given set of parameters and of $\phi_c$ is only a few GeV. If the Simplified Model really encodes the $m_s^2$ dependence of the phase transition, one expects to see some correlation between what was found numerically and what we derived analytically.

This comparison is shown in Fig.~\ref{ms2limitAll} for the parameters listed in Table~\ref{tablemsAll}, where we present a parametric plot of $m_s^2$, obtained when a first order phase transition to the physical vacuum is realized, and the upper bound derived in Eq.~(\ref{msrange}). The solutions present a clear linear correlation between the two variables. Note that the different lines appearing in Figs.~\ref{ms2limitAll} coincide for parameter sets which have the same value of $\kappa$ and $\kappa A_\kappa$ (for the same $\lambda$), which are the parameters which govern the contributions not included in the Simplified Model. A similar correlation is observed in Fig.~\ref{ms2limmh}, where we now plot the solutions and the bound corresponding to parameter given in Table~\ref{table_msmh}.

Larger values of $\kappa$ lead to smaller values of $m_s^2$, while larger negative values of $\kappa A_{\kappa}$ lead to larger values of $m_s^2$, consistent with the perturbations that these parameters perform in the $S$-dependent potential. Therefore, since the perturbations induced by $\kappa$ and the loop corrections are small, we see that consistent solutions, with a strong first order phase transition and no additional global minima at zero and finite temperature tend to be obtained close to the upper bound on $m_s^2$ derived for the Simplified Model. The values of $\phi_c/T_c$ obtained in the full model are also in close correspondence with the ones derived in the Simplified Model, for the associated values of  $m_s^2$ shown in Fig.~\ref{ms2limitAll}.

\subsection{Phase Transition Strength and the CP-Even Higgs Spectrum}

We analyze the dependence of the phase transition strength on the lightest and second lightest CP-Even Higgs masses. As we discussed before and exemplified in the Simplified Model, EWSB and baryogenesis maybe achieved in two ranges of the singlet soft susy breaking parameter, $m_s^2$. We concentrate on the small $m_s^2$ region which opens up EWSB and baroyogenesis solutions for values of the lightest CP-even Higgs mass less than about 10 GeV~(see Fig.~\ref{mh1mh2ma1mchiAll}).

\begin{figure}[!t]
\begin{center}
\includegraphics[width=0.9\textwidth,trim=0.95in 3.5in 0.95in 3.7in,clip=true]{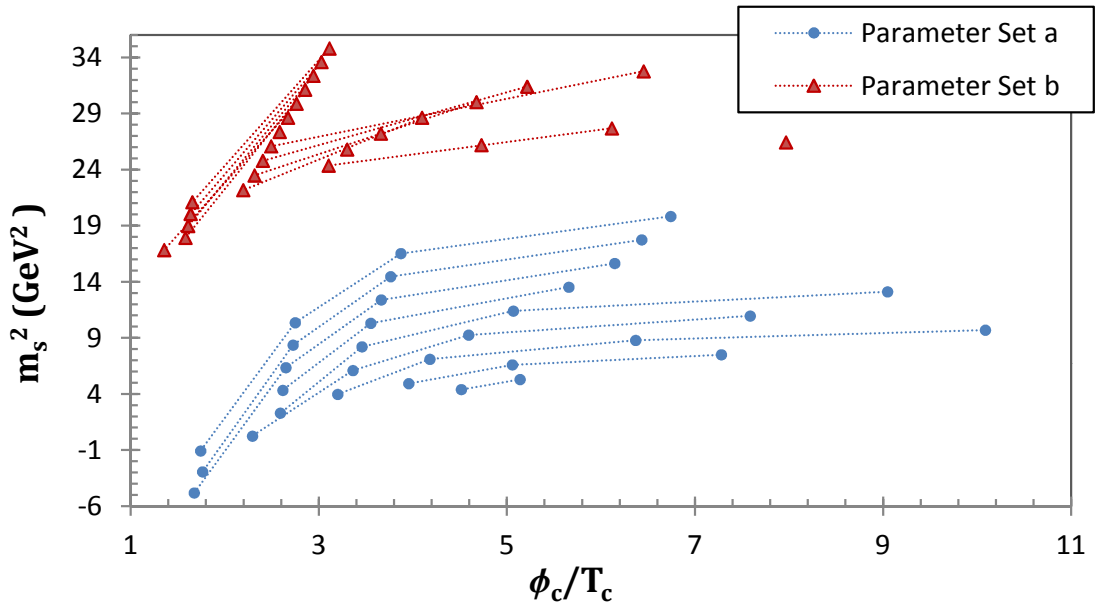}
\caption{\footnotesize{Values of $\phi_c/T_c$ vs. $m_s^2$ for Baryogenesis for both set $a$ (blue dots), and set $b$ (red triangles). Interpolating lines join points with the same values of $\Delta \lambda_{\tilde{t}}$~(and hence approximately $m_{h_2}$). $m_{h_2}$ varys in the range 115--130 GeV for Set $a$ and 100-130~GeV for Set $b$, increasing from bottom to top in intervals of about 1.5~GeV.}}
\label{phiTcms2}
\end{center}
\end{figure}

\begin{figure}[!t]
\begin{center}
\includegraphics[width=0.9\textwidth,trim=0.95in 3.5in 1in 3.7in,clip=true]{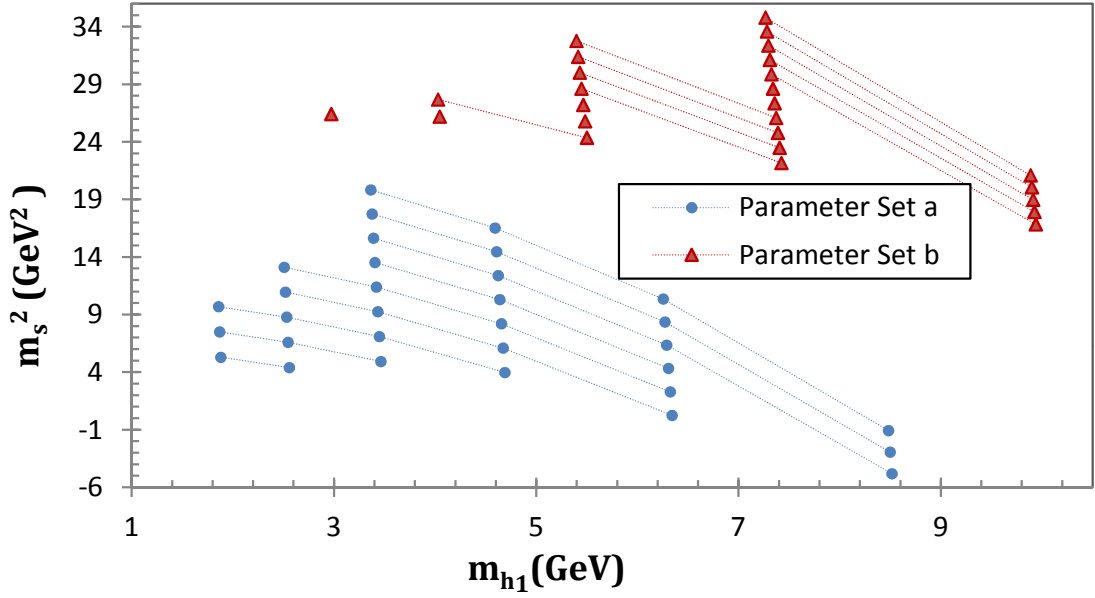}
\caption{\footnotesize{Same as Fig.~\ref{phiTcms2} but for values of $m_{h_1}$ vs. $m_s^2$ for Baryogenesis. Interpolating lines join points with the same values of $m_{h_2}$, varying from about 115~GeV to 130 GeV (bottom left to top right) for Set $a$ and from 100~GeV to 130~GeV (bottom left to top right) for Set $b$.} }
\label{mh1ms2}
\end{center}
\end{figure}

Fig.~\ref{phiTcms2} shows the dependence of $\phi_c/T_c$ on $m_s^2$, for both sets $a$~(blue dots) and $b$~(red triangles), consistent with a strong first order phase transition. We scan the parameter $\Delta \tilde{\lambda}_{\tilde{t}}$, so that $m_{h_2}$ varies in the range 100--130~GeV, and the interpolating lines join points with the same values of $m_{h_2}$. As shown in Figs.~\ref{ms2limitAll} and \ref{ms2limmh}, the behavior in the full model is similar to the one in the Simplified Model. For a given $m_{h_2}$, the values of $m_s^2$ vary in a relatively small range. As the value of $\phi_c/T_c$ increases, the critical temperature decreases smoothly, and $m_s^2$ approaches an upper bound showing a behavior similar to what would be expected in the Simplified Model as can be seen in Fig.~\ref{simplified1}.

As stressed before, larger values of $m_s^2$ correspond to smaller values of $S_0$, which in turn lead to smaller values of $m_{h_1}$,  Eq.~(\ref{mh1}), in the region of parameters under study. Fig.~\ref{mh1ms2} shows the dependence of $m_{h_1}$ on $m_s^2$ for sets $a$~(blue dots) and $b$~(red triangles) and the interpolated lines are associated with fixed values of $m_{h_2}$ as in Fig.~{\ref{phiTcms2}. As anticipated, for fixed values of $m_{h_2}$, Fig.~\ref{mh1ms2} shows the inverse correlation between $m_{h_1}$ and $m_s^2$.

\begin{figure}[!b]
\begin{center}
\includegraphics[width=0.9\textwidth,trim=0.9in 3.5in 1.3in 3.85in,clip=true]{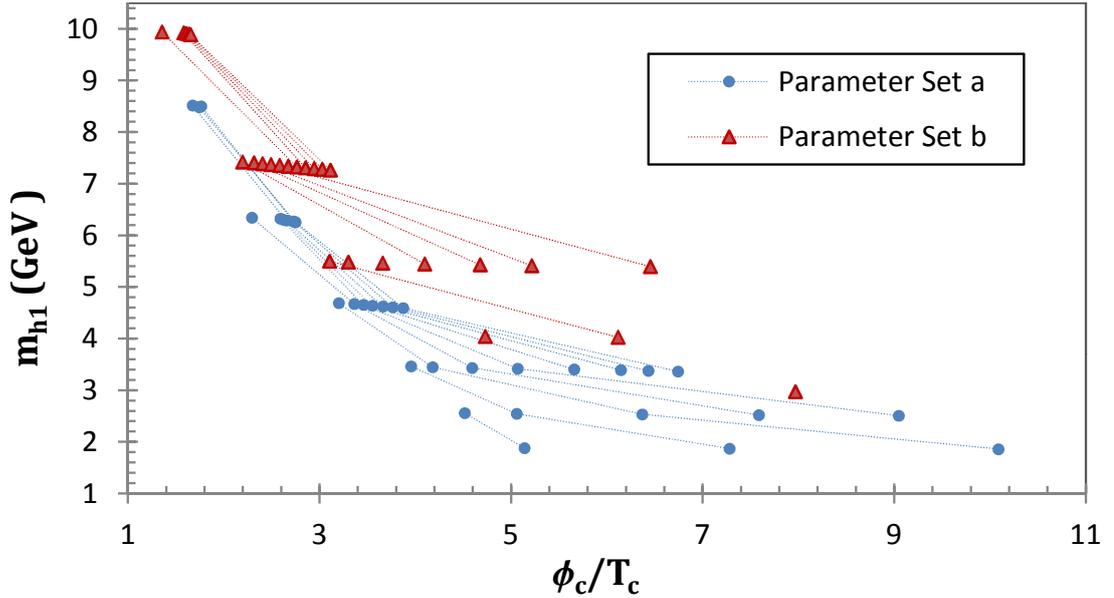}
\caption{\footnotesize{Same as Fig.~\ref{phiTcms2}, but for the values of $\phi_c/T_c$ as a function of $m_{h_1}$. }}
\label{phiTcmh1}
\end{center}
\end{figure}

Fig.~\ref{phiTcmh1} shows the dependence  of $\phi_c/T_c$ on the lightest CP-even Higgs mass, showing that indeed the lighter the CP-even Higgs boson is, the stronger the phase transition becomes, leading to very interesting phenomenological properties, which we shall discuss in more detail in the next section. From Figs.\ref{phiTcmh1} and \ref{mh1ms2} we see  that for the same value of $m_{h_1}$, the phase transition becomes stronger for larger values of $m_{h_2}$, which are associated with larger values of $m_s^2$.

\subsection{Transition Temperature}

\begin{figure}[!t]
\vspace{1cm}
\centering
\includegraphics[width=0.85\textwidth]{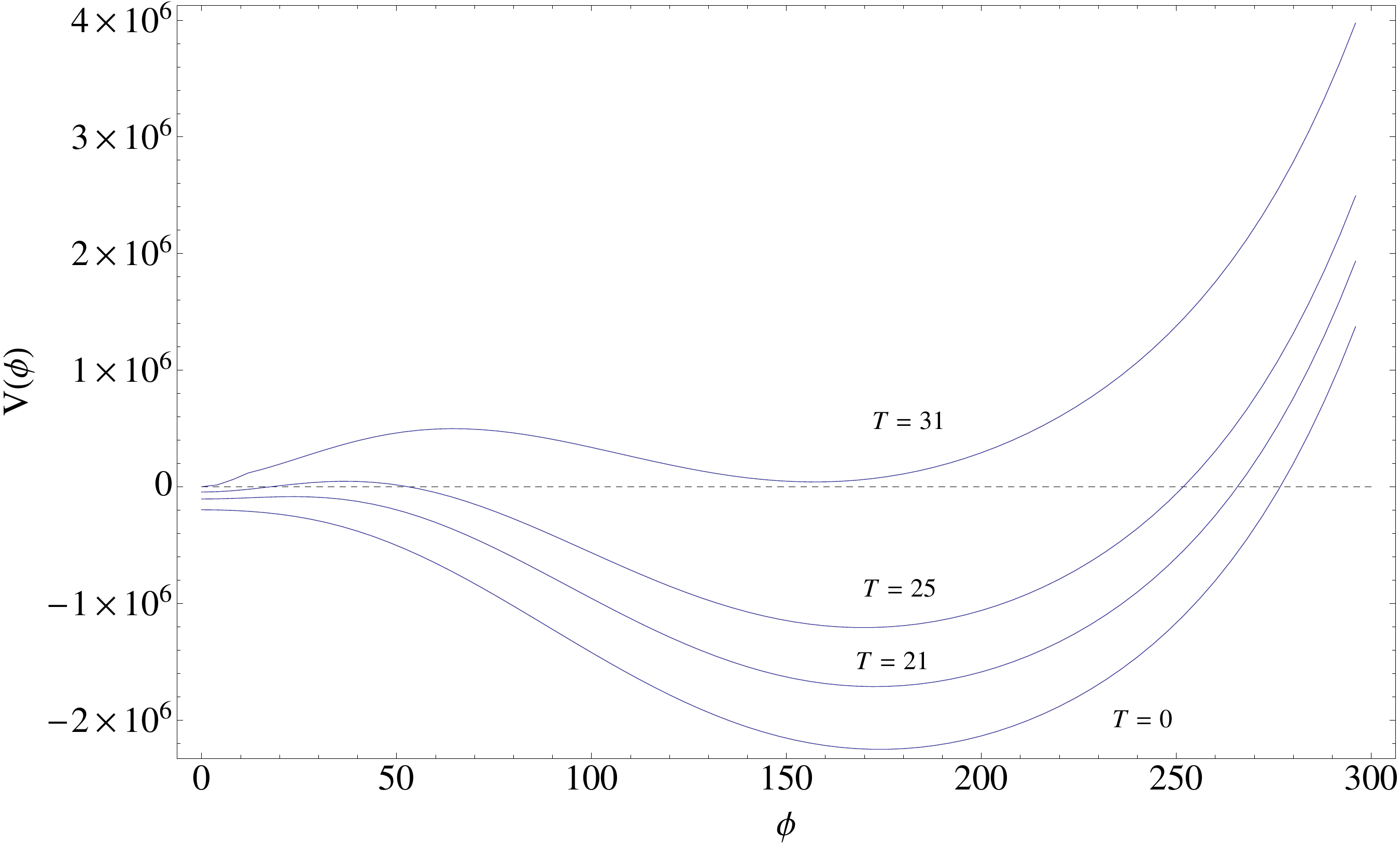}
\vspace{1cm}
\caption{\footnotesize{Potential showing strong first-order phase transition at $T_c$ and electroweak symmetry breaking at $T=0$ for Parameter Set $a$ given in Table~\ref{table_msmh} with $m_{h_2}\sim 115$ GeV,  and $m_{h_1}\sim 1.8$ GeV. The potential is shown along the minimal energy barrier path, as explained in the text. Temperatures are given in GeV. }}
\label{phicTcPot}
\vspace{1cm}
\end{figure}
The tunneling probability per unit time and unit volume from the false~(symmetric) to the real~(broken) minimum in a thermal bath is given by~\cite{Linde:1981zj}
\begin{equation}
\frac{\Gamma}{\nu}\sim A(T)\exp{\left[-\frac{S_3(T)}{T}\right]}\;,
\label{bounce3}
\end{equation}
where the prefactor is $A(T)\simeq T^4$ and $S_3$ is the three-dimensional effective action. At high temperature the euclidean action simplifies to:
\begin{equation}
S_3=4\pi\int_0^\infty r^2 dr \left[\frac{1}{2} \left(\frac{d \phi}{dr}\right)^2\;
+V(\phi,T)\right]
\end{equation}
where $r^2=\vec x^2$. The euclidean equations of motion yield
\begin{equation}
\frac{d^2\phi}{dr^2}+\frac{2}{r}\frac{d\phi}{dr}=V^\prime(\phi,T)\;,
\end{equation}
for the bounce solution, with boundary conditions $\lim_{r\to\infty}\phi(r)=0$ and $\left.d\phi/dr\right|_{r=0}=0$.

The nucleation temperature $T^n$ is defined as the temperature at which the probability for a bubble to be nucleated inside a horizon volume is of order one. Below $T^n$ the transition continues until a temperature $T^t$ when the fraction of the causal horizon in the broken phase is of order one~\cite{Anderson:1991zb,thomas}.

In Fig.~\ref{phicTcPot} we plot the effective potential for $\phi$ for parameter set $a$ with $m_{h_1} \simeq 1.8$~GeV and $m_{h_2} \simeq 115$~GeV for a series of different temperatures from zero to the critical temperature. We chose this as an example since as we will show in the next section, such low values of $m_{h_1}$ are correlated with large spin-independent direct DM detection cross-sections, as the ones suggested by the CoGeNT and DAMA experiments. Additionally, as shown in Fig.~\ref{phiTcmh1}, large values of $\phi_c/T_c$ correspond to the smallest values of $m_{h_1}$. The non-trivial minimum, $\phi_0(T)$, of the potential is always less than or equal to $v$ and hence small values of $m_{h_1}$, for which there is a strong first order phase transition, are associated with relatively small values of $T_c$.

The potential in Fig.~\ref{phicTcPot} is plotted along the minimal energy barrier path between the trivial and nontrivial minima of $\phi$: For each value of $\phi$ we found the values of $S$ and $\tan\beta$ that minimize the potential. This defines a smooth path along $H_u$, $H_d$ and $S$ that connects both minima. It is important to stress that the barrier between both minima slowly disappears as $T$ goes below $T_c$. The nucleation temperature is always a few GeV below the critical temperature. Note that the potential is normalized such that $V(H_u=0,H_d=0,S=0)=0$.

For transition temperatures of the order of the weak scale, nucleation occurs when $S_3(T^n)/T^n\sim 135$ continuing until $S_3(T^t)/T^t\sim 110$~\cite{Anderson:1991zb,thomas}. Hence, even though for our particular example the critical temperature is about 31~GeV, the transition does not end until $T ^t \simeq 21$~GeV~(the nucleation temperature is about 22~GeV). As this examples shows, for small values of $m_{h_1}$, the phase transition strength, parameterized by $\phi_0(T^t)/T^t$ (or $\phi_0(T^n)/T^n$), may be even larger than what the value of $\phi_c/T_c$ indicates. Such strong phase transitions make the possible effect of magnetic fields on the sphaleron energy (and hence on the final baryon asymmetry)~\cite{DeSimone:2011ek} much less relevant than in the MSSM.

For values of $m_{h_1}$ larger than the one considered above, the first order phase transition becomes weaker, and the difference between the critical temperature and the nucleation temperature becomes smaller. Still, as can be seen in Fig.~\ref{phiTcmh1} even for values of $m_{h_1}$ as large as 5~GeV, the values of $\phi_c/T_c \simgt 3$ and the phase transition is strong enough to preserve the baryon asymmetry, even after possible magnetic field effects. On the other hand, small values of $m_{h_1}$, leading to $\phi_c/T_c \simgt 10$, may lead to very small transition temperatures, below the DM freeze out temperature or the nucleosynthesis temperature. In fact, for small enough $m_{h_1}$, there is a possibility of a large barrier even at zero temperature, giving rise to the situation where there is no transition from the false vacuum.  Hence, although solutions with $\phi_c/T_c \simgt 10$ may be found, we have not considered such a possibility.

\section{Phenomenological Consequences}

In general, the phenomenology of the region of parameters under study is very similar to the one analyzed in Ref.~\cite{Draper:2010ew}. The presence of light CP-even and CP-odd Higgs bosons tend to be strongly constrained by the LEP experiment. In particular, the possible associated production $Z \to A \; H$ has been studied in detail~\cite{Bechtle:2004ig}. This production channel demands $A$ and $H$ to have relevant weak couplings. Near the PQ symmetry limit, however, the non-standard MSSM-like CP-even and CP-odd Higgs bosons become very heavy and the light non-standard CP-even and CP-odd Higgs bosons, $h_1$ and $a_1$, are mostly singlet-like, leading to a very strong suppression of this production cross section, and therefore to weak bounds from this channel. In addition, the decays of the SM-like Higgs boson, $h_2$, to $h_1h_1$ and $a_1a_1$ pairs are generically suppressed~\cite{Draper:2010ew}. Thus $h_1$ and $a_1$ are hidden from four-fermion searches at both LEP~\cite{Schael} and the Tevatron~\cite{Abazov:2009yi} designed to test a light $a_1$ scenario. This also implies that, unless other light particles appear in the spectrum, the bounds on the SM-like Higgs boson mass are similar to the SM Higgs bounds. However, as mentioned before, lower SM-like Higgs masses make it easier to find stable solutions with small values of $m_{h_1}$ as shown in Fig.~\ref{mh1mh2ab}. Hence, in our work we also considered values of $m_{h_2}$ below the LEP SM-Higgs limits that could be avoided if additional non-standard decays were present.

In this model, additional decay modes of the SM-like Higgs boson may appear, for instance, for low values of the bino mass. In such a case the SM-like Higgs boson may decay in the following way~\cite{LiuZhang}
\begin{equation}
h_2 \to \chi_2 \chi_1 \to h_1 \chi_1 \chi_1  \; (f \bar{f} + {\rm Miss. Energy})\;,
\label{h2toh11}
\end{equation}
or, for sufficiently low values of $m_{\chi_2}$,
\begin{equation}
h_2 \to \chi_2 \chi_2 \to h_1 h_1 \chi_1 \chi_1 \; (2 \times f \bar{f} + {\rm Miss. Energy})\;.
\label{h2toh12}
\end{equation}
This leads to a decay into missing energy and somewhat soft jets or leptons, for which the limits on the SM-like Higgs may be relaxed~\cite{Chang:2007de}. Specifically, bounds on the SM-like Higgs mass coming from the search of Higgs bosons decaying into bottom quark pairs may be lowered if the branching ratio of its decay into bottom quarks is about 20 \% or smaller~\cite{Barate:2003sz}.  A specific analysis of LEP data for the decay channels described in Eqs.~(\ref{h2toh11}) and (\ref{h2toh12})  is not available, and should be performed to determine the viability of this region of parameter space. 

We used NMSSMTools~\cite{Ellwanger:2004xm} to verify our zero temperature results. For the same parameter values we used, this program requires somewhat different value of $\mu$~($S_0$) due to a slightly different treatment of the zero temperature radiative corrections to the effective potential. For the same value of the $m_{h_1}$ mass, however, the rest of the spectrum was at most $10\%$ away from the values we obtained. We verified that the branching fraction into bottom quarks may be lowered to $\sim 20\%$ if, for example, the hypercharge gaugino mass is of the order 40--50~GeV.  Hence, values of $m_{h_2} \simeq 100$~GeV become consistent with the LEP bound on the $H \to b \bar{b}$ channel. Additionally, we checked that all other phenomenological constraints were fulfilled, including all rare B-decay constraints. The only exception was the  relic density, which is very sensitive to the exact spectrum due to its dependence on near-resonant annihilation, as we shall discuss in detail below. 

It is worth mentioning that  the non-standard decay channels can also occur for SM-like Higgs masses larger than the LEP bound. Their main effect is to decrease the branching ratio of the SM-like Higgs boson into standard model particles, including the decays into photons, bottom quarks and $W$ bosons, which constitute the main search channels for a light SM-like Higgs at the LHC and the Tevatron. Therefore, the possibility of extra decay modes will be tested at these colliders, as they also expand the Higgs searches to masses below the LEP bounds on the SM Higgs mass.

The spin independent direct dark matter detection cross section is approximately given by~\cite{Draper:2010ew}
\begin{align}
\sigma_{\rm SI} \approx
\frac{\left (\left(\frac{\varepsilon}{0.04} \right)+ 0.46 \left (\frac{\lambda}{0.1} \right) \left(\frac{v}{\mu}\right) \right)^2 \left(\frac{y_{h_1\chi_1\chi_1}}{0.003}\right)^2 10^{-40} \mathrm{cm}^2}{ \left(\frac{m_{h_1}}{1 \mathrm{GeV}}\right)^4 },
\end{align}
where the $h_1 \chi_1 \chi_1$ coupling is $y_{h_1 \chi_1 \chi_1} \approx -\sqrt{2} \kappa$ for a singlino-like $\chi_1$ and a singlet-like $h_1$.  The Higgsino mass parameter $\mu$ is of the order 185~GeV for set $a$ and 215~GeV for set $b$.

\begin{figure}[!t]
\begin{center}
\includegraphics[width=0.9\textwidth,trim=1in 3.5in 1.3in 3.85in,clip=true]{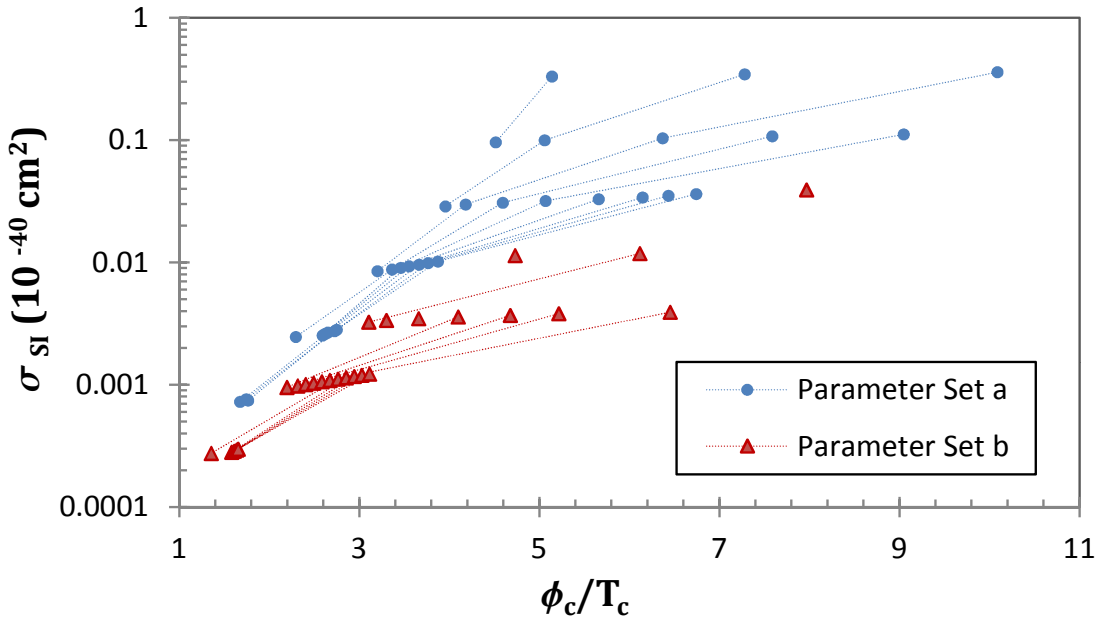}
\caption{\footnotesize{Same as Fig.~\ref{phiTcms2}, but showing the correlation between $\phi_c/T_c$ and the spin independent cross section. Lower values of $m_{h_2}$ appear on top in this Figure. }}
\label{phicTcsig}
\end{center}
\end{figure}

The spin independent direct dark matter detection cross section increases with the fourth power of $m_{h_1}^{-1}$. Additionally, as we showed in the previous section, small values of $m_{h_1}$ tend to be associated with a strong first order phase transition. These two observations lead us to note an interesting correlation between a large spin independent direct detection dark matter cross section and a strong first order phase transition. This is shown in Fig.~\ref{phicTcsig}.

\begin{figure}[!t]
\begin{center}
\includegraphics[width=0.9\textwidth,trim=.8in 3.4in 0.85in 3.7in,clip=true]{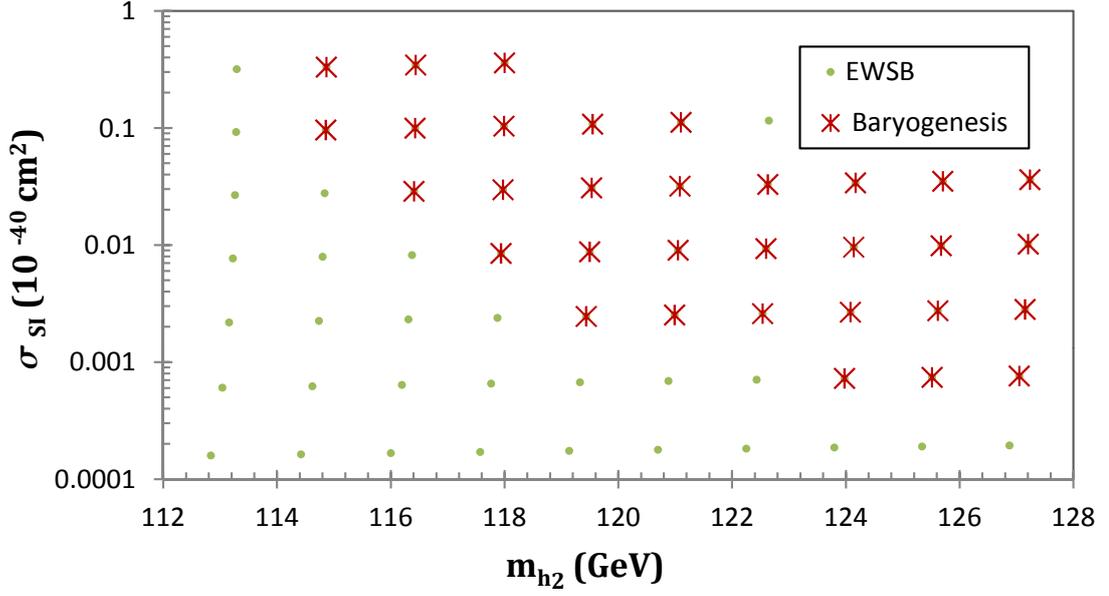}
\caption{\footnotesize{Values of $m_{h2}$ vs. $\sigma_{\rm{SI}} (10^{-40} \rm{cm}^{2})$ for both EWSB (Green dots) and Baryogenesis (Red crosses) for Parameter Set $a$. }}
\label{mh2siga}
\end{center}
\end{figure}

As shown in the previous section, very small values of $m_{h_1}$ in sets $a$ and $b$ are obtained for values of $m_{h_2}$ close to the LEP SM Higgs bound. For parameter set $a$, as shown in Fig.~\ref{mh2siga}, values of the spin independent direct dark matter detection cross section as large as the ones consistent with potential DM signatures observed by CoGeNT and DAMA~\cite{Bernabei:2008yi,Aalseth:2010vx} may be obtained for values of $m_{h_2}  \simlt 120$~GeV. Instead, for the parameter set $b$, as seen in Fig.~\ref{mh2sigb}, the direct detection cross section tends to be an order of magnitude smaller than for Set $a$ and only approaches values necessary to explain these signatures when $m_{h_2}\sim 100$ GeV. For completeness in Fig.~\ref{mchi1sigab} we show the neturalino mass vs. the spin independent cross-section for parameter sets $a$ and $b$. In Figs.~\ref{mh2siga}--{mchi1sigab}, we denote with red crosses the points consistent with Baryogenesis and with green dots those that allow for EWSB but would not lead to the preservation of the baryon asymmetry at the electroweak phase transition.

\begin{figure}[!b]
\begin{center}
\includegraphics[width=0.9\textwidth,trim=.8in 3.3in 1in 3.75in,clip=true]{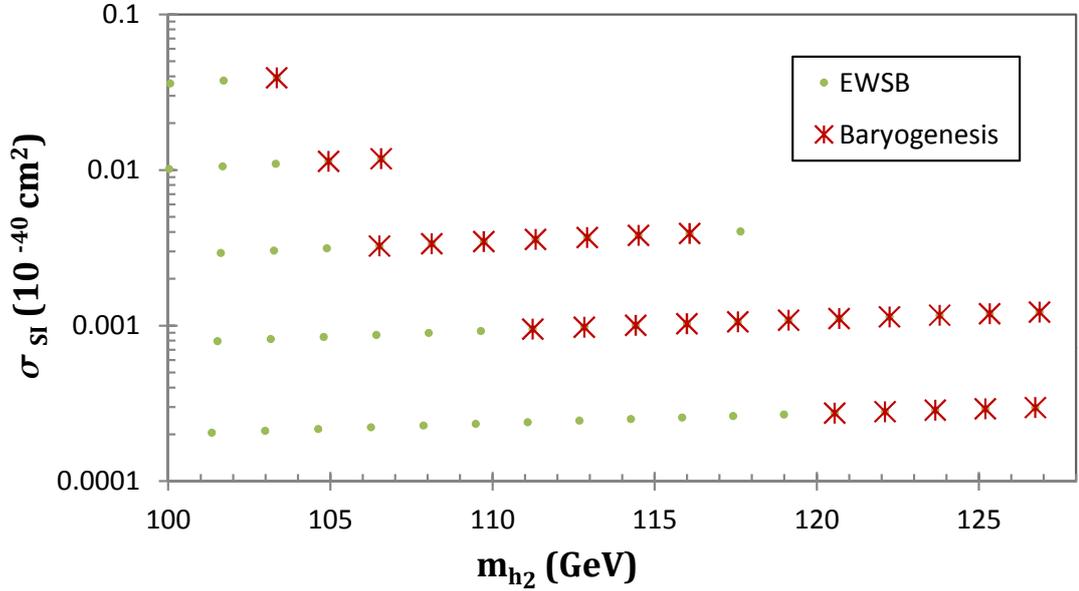}
\caption{Values of $m_{h2}$ vs. $\sigma_{\rm{SI}} (10^{-40} \rm{cm}^{2})$ for both EWSB (Green dots) and Baryogenesis (Red crosses) for Parameter Set $b$. }
\label{mh2sigb}
\end{center}
\end{figure}

\begin{figure}[!t]
\begin{minipage}[b]{0.48\linewidth}
\centering
\includegraphics[width=\textwidth,trim=.8in 3.5in 1in 3.7in,clip=true]{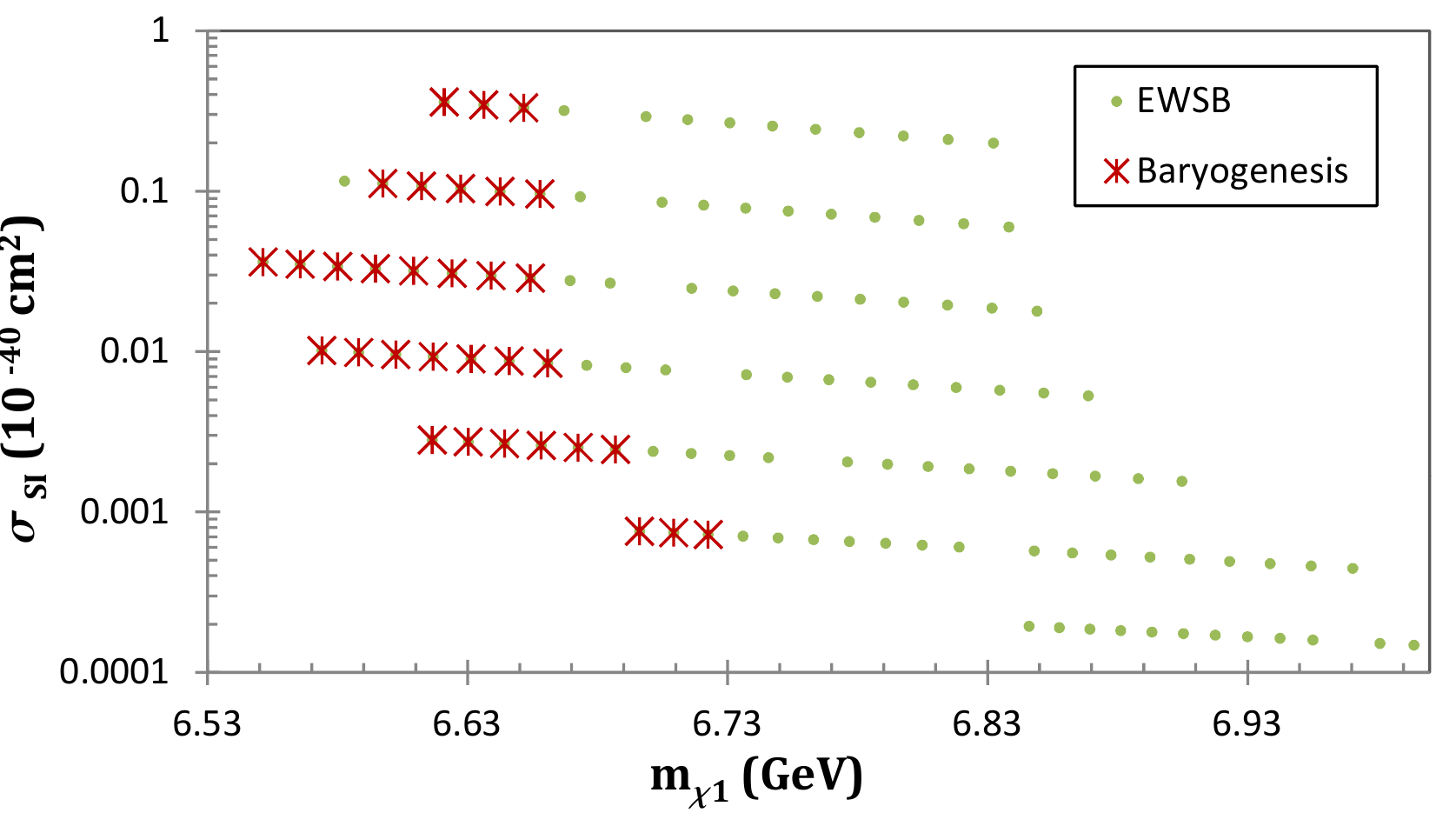}
\end{minipage}
\hspace{0.2cm}
\begin{minipage}[b]{0.48\linewidth}
\centering
\includegraphics[width=\textwidth,trim=.75in 3.5in 1in 3.7in,clip=true]{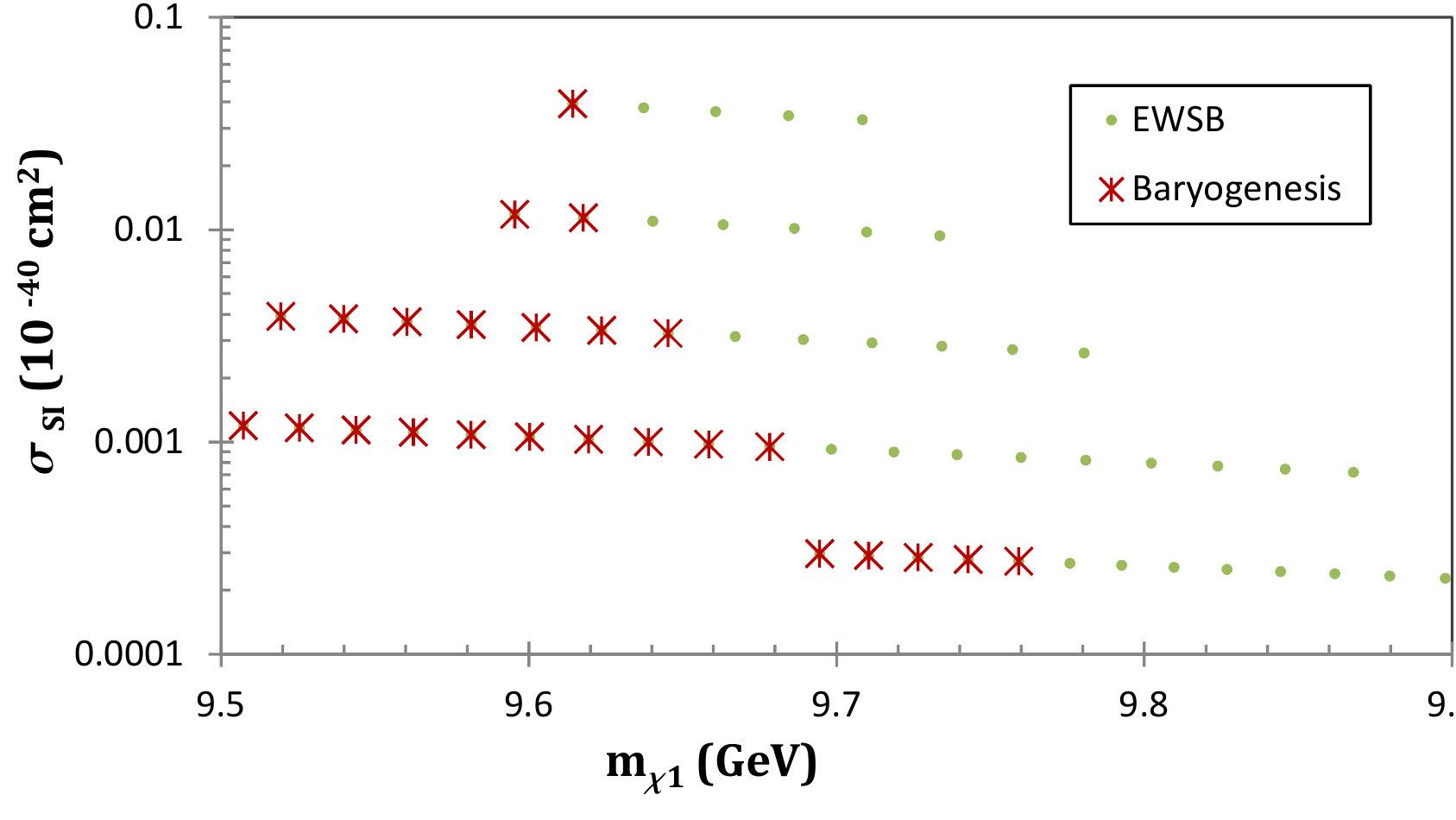}
\end{minipage}
\vspace{0.4cm}
\caption{\footnotesize{$m_{\chi_1}$ vs. $\sigma_{SI}$ for both EWSB (Green dots) and Baryogenesis (Red crosses) for Parameter Set $a$ (left) and $b$ (right). }}
\label{mchi1sigab}
\end{figure}

Regarding the Dark Matter relic density, in the region under study it is obtained by close to resonance annihilation at finite temperature via a light CP-odd Higgs boson~\cite{Draper:2010ew},
\begin{eqnarray}
\Omega h^2 \approx \frac{0.1\left(\frac{m_{a_1}}{15{\rm GeV}} \right) \left(\frac{\Gamma_{a_1}}{10^{-5} {\rm GeV}} \right) \left(\frac {\mu} {v}\right)^2\left(\frac{0.003}{y_{a_1\chi_1\chi_1}}\right)^2 \left(\frac{0.1}{\lambda}\right)^2 } {{\rm erfc}\left(\frac{2m_{\chi_1}}{m_{a_1}}\sqrt{x_f |1 - m_{a_1}^2/(4 m_{\chi_1}^2)|}\right)/{\rm erfc}\left(2.2\right)} \label{rd}
\end{eqnarray}
where $x_f=m_{\chi_1}/T_f$ is the freeze-out point, $\Gamma_{a_1}$ is the width of $a_1$ and $y_{a_1 \chi_1 \chi_1} \simeq \sqrt{2} \kappa$. Since at zero temperature the annihilation cross section is off-resonance~($m_{\chi_1}\sim 6.5$~GeV and $m_{a_1}\sim 15.5$~GeV for set $a$ and $m_{\chi_1}\sim 9.5$~GeV and $m_{a_1}\sim 22$~GeV for set $b$), at current times it is much smaller than $10^{-36}$~cm$^2$, and hence the bounds coming from the modification of the antiproton and gamma ray fluxes~\cite{Adriani} become very weak in this model.

Finally, Fig.~\ref{sigOmab} shows the correlation of the spin independent cross section with the obtained Dark Matter relic density for both sets $a$ and $b$. Observe that in the parameter sets $a$ and $b$ the relic density is slightly above the measured value. We have not attempted to tune these values, but given that the relic density is obtained from a resonant condition a small variation of the parameters would adjust its value to the observed one.  For instance, a small increase of $\kappa$ by less than one percent would be enough to bring the relic density to agreement with observations, without modifying any other relevant phenomenological property.
\begin{figure}[!b]
\begin{minipage}[b]{0.48\linewidth}
\centering
\includegraphics[width=\textwidth,trim=.9in 3.5in 1in 3.7in,clip=true]{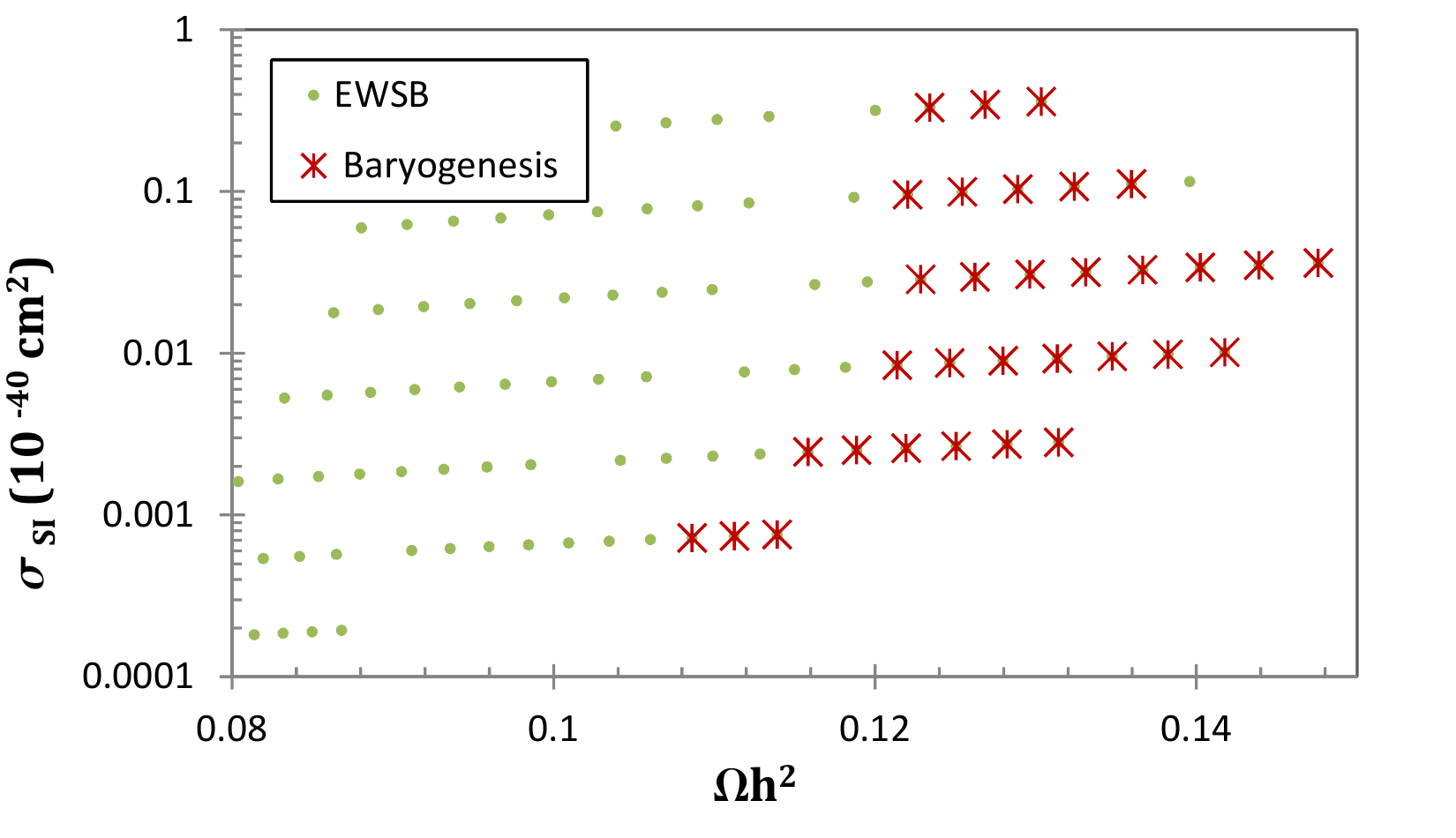}
\end{minipage}
\hspace{0.2cm}
\begin{minipage}[b]{0.48\linewidth}
\centering
\includegraphics[width=\textwidth,trim=.9in 3.5in 1in 3.7in,clip=true]{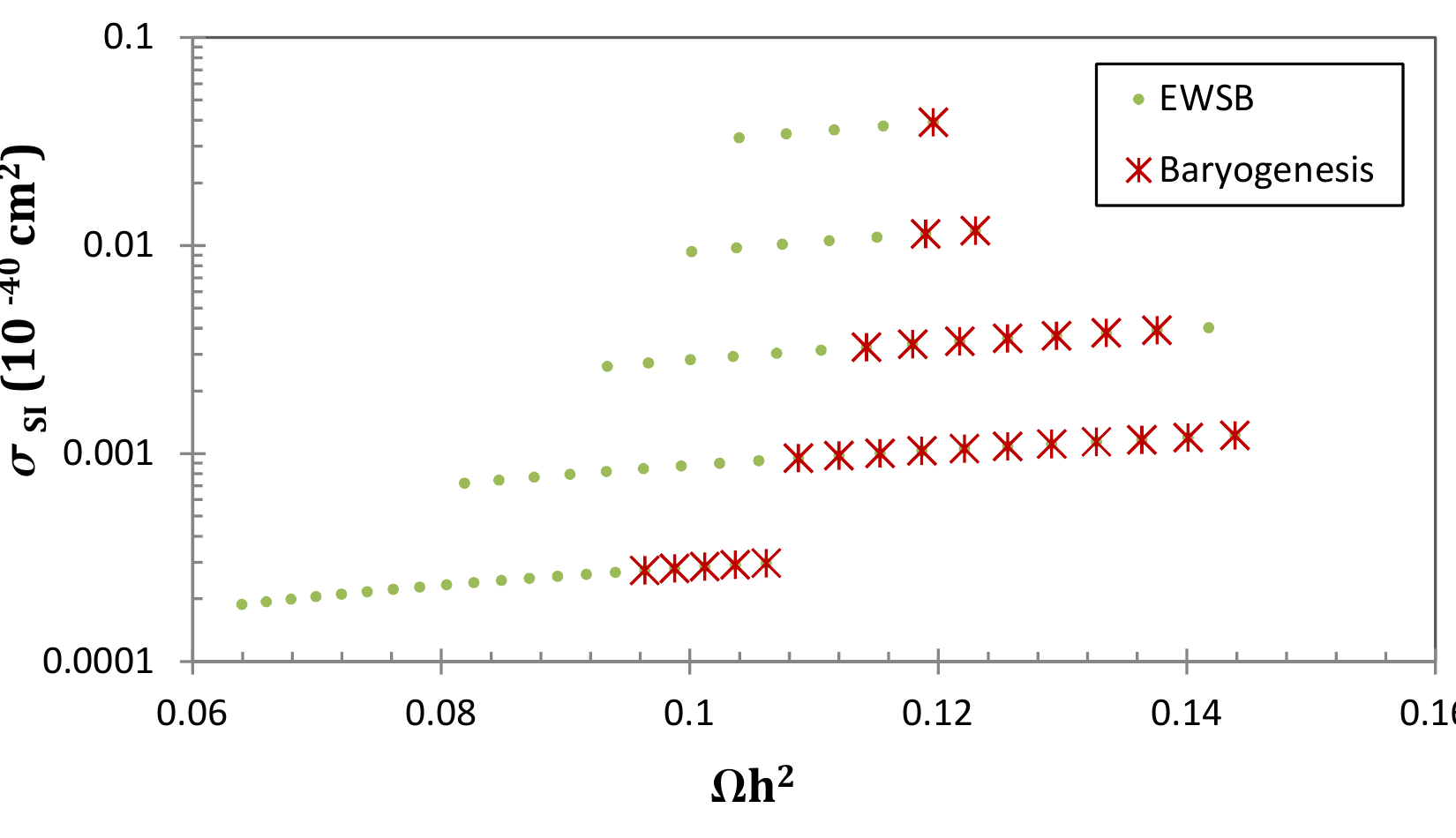}
\end{minipage}
\vspace{0.2cm}
\caption{\footnotesize{$\Omega h^2$ vs. $\sigma_{SI}$ for both EWSB (Green dots) and Baryogenesis (Red crosses) for Parameter Set $a$ (left) and $b$ (right). }}
\label{sigOmab}
\end{figure}

\section{Conclusions}

In this article we have examined the electroweak phase transition in the NMSSM close to the PQ symmetry limit. This model is characterized by a spectrum of light CP-even and CP-odd Higgs particles, as well as a light neutralino. In general, $h_1$ masses below 10~GeV, and light CP-odd Higgs bosons and neutralino masses below 25~GeV are obtained in the region consistent with a strong first order phase transition. This leads to interesting phenomenological consequences for collider and direct dark matter detection experiments.

Although the neutralinos are predominantly singlinos, an experimentally consistent relic density may be obtained if the neutralinos annihilate resonantly via the interchange of the light scalars. Additionally, the light CP-even scalars may lead to a large direct dark matter detection cross section, consistent with the recent observations at the CoGeNT, DAMA~(and CRESST) experiments.

It is of interest to study if this model can also lead to the generation of the baryon asymmetry at the electroweak scale. For this purpose, we studied the electroweak phase transition properties.  We found that the electroweak phase transition tends to be first order and its strength is enhanced for small values of the lightest CP-even Higgs mass, establishing a correlation between the strength of the first order phase transition and the size of the direct dark matter detection cross section.

In our numerical study we found regions of parameter space where a strong first order phase transition occurs yielding the proper relic density. The lightest neutralino mass is in the 6--10 GeV range and the lightest Higgs mass, $m_{h_1}$, is below a few GeV, thus enhancing the cross section to values close to the ones required for an explanation of the CoGeNT and DAMA experimental data. The mass of the SM-like Higgs boson, $m_{h_2}$, is in the low mass window compatible with LEP and Tevatron/LHC data. There are also regions of space in which the SM-like Higgs mass is slightly below the LEP SM Higgs bound. However, this bound may be avoided due to non-standard Higgs decays into neutralinos. We expect that a more computationally intensive scan would slightly enlarge the LSP mass window consistent with a strong first order phase transition and a large direct DM detection cross-section.

~\\

{\large{\textbf{Acknowledgements:}}}
\vspace{0.2 cm}

Fermilab is operated by Fermi Research Alliance, LLC under Contract No. DE-AC02-07CH11359 with the U.S. Department of Energy. Work at ANL is supported in part by the U.S. Department of Energy~(DOE), Div.~of HEP, Contract DE-AC02-06CH11357. This work was supported in part by the DOE under Task TeV of contract DE-FGO2-96-ER40956. M.~C. and C.~W. would like to thank the Aspen Center for Physics, where part of this work has been done.

\end{document}